\documentclass[iop]{emulateapj}

\usepackage{color,natbib,multirow}

\shorttitle{\ion{He}{2}~$\lambda$4686 in $\eta$~Car}
\shortauthors{Teodoro et al.}

\begin{document}

\newcommand{\heii}{\ion{He}{2}~$\lambda4686$}
\newcommand{\ec}{$\eta$~Car}

\title{\ion{H\MakeLowercase{e}}{2}~$\lambda4686$ in $\eta$~Carinae: collapse of the wind-wind collision region during periastron passage}

\author{M.~Teodoro}
\author{A.~Damineli}
\affil{Instituto de Astronomia, Geof\'{\i}sica e Ci\^encias Atmosf\'ericas, Universidade de S\~ao Paulo,\\ Rua do Mat\~ao 1226, Cidade Universit\'aria, S\~ao Paulo, 05508-900, Brazil}

\author{J.~I.~Arias} 
\affil{Departamento de F\'{\i}sica, Universidad de La Serena, Av. Cisternas 1200 Norte, La Serena, Chile}
\author{F.~X.~de Ara\'ujo\altaffilmark{1}}
\affil{Observat\'orio Nacional, Rua General Jos\'e Cristino 77, S\~ao Crist\'ov\~ao, Rio de Janeiro, 20921-400, Brazil}
\author{R.~H.~Barb\'a\altaffilmark{2}} 
\affil{Instituto de Ciencias Astron\'omicas, de la Tierra, y del Espacio (ICATE-CONICET), Av. Espa\~na Sur 1512, J5402DSP San Juan, Argentina}
\author{M.~F.~Corcoran\altaffilmark{3}} %
\affil{CRESST and X-ray Astrophysics Laboratory, NASA/Goddard Space Flight Center, Greenbelt, MD 20771, USA}
\author{M.~Borges~Fernandes} 
\affil{Observat\'orio Nacional, Rua General Jos\'e Cristino 77, S\~ao Crist\'ov\~ao, Rio de Janeiro, 20921-400, Brazil}
\author{E.~Fern\'andez-Laj\'us\altaffilmark{4}} 
\affil{Facultad de Ciencias Astron\'omicas y Geof\'{\i}sicas, Universidad Nacional de La Plata, Paseo del Bosque s/n, La Plata, BA, B1900FWA, Argentina}
\author{L.~Fraga}
\affil{Southern Observatory for Astrophysical Research, Colina El Pino s/n, Casilla 603, La Serena, Chile}
\author{R.~C.~Gamen\altaffilmark{4}} 
\affil{Facultad de Ciencias Astron\'omicas y Geof\'{\i}sicas, Universidad Nacional de La Plata, Paseo del Bosque s/n, La Plata, BA, B1900FWA, Argentina}
\author{J.~F.~Gonz\'alez}
\affil{Instituto de Ciencias Astron\'omicas de la Tierra y del Espacio (ICATE-CONICET), Avenida Espa\~na 1512 Sur, J5402DSP, San Juan, Argentina}
\author{J.~H.~Groh}
\affil{Max-Planck-Institute f\"ur Radioastronomie, Auf dem H\"ugel 69, D-53121 Bonn, Germany}
\author{J.~L.~Marshall}
\affil{Department of Physics and Astronomy, Texas A\&M University, College Station, TX 77843-4242, USA}
\author{P.~J.~McGregor}
\affil{Research School of Astronomy and Astrophysics (RSAA), Mount Stromlo Observatory, Cotter Road, Weston, ACT 2611, Australia}
\author{N.~Morrell} 
\affil{Las Campanas Observatory, Observatories of the Carnegie Institution of Washington, Casilla 601, La Serena, Chile}
\author{D.~C.~Nicholls}
\affil{Research School of Astronomy and Astrophysics (RSAA), Mount Stromlo Observatory, Cotter Road, Weston, ACT 2611, Australia}
\author{E.~R.~Parkin}
\affil{Research School of Astronomy and Astrophysics (RSAA), Mount Stromlo Observatory, Cotter Road, Weston, ACT 2611, Australia}
\author{C.~B.~Pereira} 
\affil{Observat\'orio Nacional, Rua General Jos\'e Cristino 77, S\~ao Crist\'ov\~ao, Rio de Janeiro, 20921-400, Brazil}
\author{M.~M.~Phillips}
\affil{Las Campanas Observatory, Observatories of the Carnegie Institution of Washington, Casilla 601, La Serena, Chile}
\author{G.~R.~Solivella\altaffilmark{4}}
\affil{Facultad de Ciencias Astron\'omicas y Geof\'{\i}sicas, Universidad Nacional de La Plata, Paseo del Bosque s/n, La Plata, BA, B1900FWA, Argentina}
\author{J.~E.~Steiner}
\affil{Instituto de Astronomia, Geof\'{\i}sica e Ci\^encias Atmosf\'ericas, Universidade de S\~ao Paulo,\\ Rua do Mat\~ao 1226, Cidade Universit\'aria, S\~ao Paulo, 05508-900, Brazil}
\author{M.~Stritzinger\altaffilmark{5}}
\affil{The Oskar Klein Centre, Department of Astronomy, Stockholm University, AlbaNova, 10691 Stockholm, Sweden}
\author{I.~Thompson} 
\affil{Observatories of the Carnegie Institution, 813 Santa Barbara Street, Pasadena, California 91101, USA}
\author{C.~A.~O.~Torres} 
\affil{Laborat\'orio Nacional de Astrof\'{\i}sica, Rua Estados Unidos 154, Bairro das Na\c c\~oes, 37504-364, Itajub\'a, Brazil}
\author{M.~A.~P.~Torres\altaffilmark{6}}
\affil{Harvard-Smithsonian Center for Astrophysics, 60 Garden Street, Cambridge, MA 02138, USA} 
\author{M.~I.~Zevallos~Herencia}
\affil{Observat\'orio Nacional, Rua General Jos\'e Cristino 77, S\~ao Crist\'ov\~ao, Rio de Janeiro, 20921-400, Brazil}

\altaffiltext{1}{\emph{in memoriam}}
\altaffiltext{2}{Departamento de F\'{\i}sica, Universidad de La Serena, Cisternas 1200 Norte, La Serena, Chile}
\altaffiltext{3}{Universities Space Research Association, 10211 Wincopin Circle, Columbia, MD USA 21044}
\altaffiltext{4}{Instituto de Astrof\'{\i}sica de La Plata -- CONICET}

\altaffiltext{5}{Dark Cosmology Centre, Niels Bohr Institute, University of Copenhagen, Juliane Maries Vej 30, 2100 Copenhagen \O, Denmark}
\altaffiltext{6}{SRON, Netherlands Institute for Space Research, 3584 CA, Utrecht, the Netherlands}

\email{e-mail to: mairan@astro.iag.usp.br}

\begin{abstract}

The periodic spectroscopic events in $\eta$~Carinae are now well established and occur near the periastron passage of two massive stars in a very eccentric orbit. Several mechanisms have been proposed to explain the variations of different spectral features, such as an eclipse by the wind-wind collision boundary, a shell ejection from the primary star or accretion of its wind onto the secondary.  All of them have problems explaining all the observed phenomena. To better understand the nature of the cyclic events, we performed a dense monitoring of $\eta$~Carinae with 5 Southern telescopes during the 2009 low excitation event, resulting in a set of data of unprecedented quality and sampling.  The intrinsic luminosity of the \ion{He}{2}~$\lambda4686$ emission line ($L\sim310~L_\odot$) just before periastron reveals the presence of a very luminous transient source of extreme UV radiation emitted in the wind-wind collision (WWC) region. Clumps in the primary's wind probably explain the flare-like behavior of both the X-ray and \ion{He}{2}~$\lambda4686$ light-curves.  After a short-lived minimum, \ion{He}{2}~$\lambda4686$ emission rises again to a new maximum, when X-rays are still absent or very weak. We interpret this as a collapse of the WWC onto the ``surface" of the secondary star, switching off the hard X-ray source and diminishing the WWC shock cone. The recovery from this state is controlled by the momentum balance between the secondary's wind and the clumps in the primary's wind.
 
\end{abstract}


\section{Introduction}
The stellar object $\eta$~Carinae (HD~93308; hereafter $\eta$~Car) is one of the most luminous and massive of our Galaxy. Since its Great Eruption in 1843 \citep{Smith:2011p2009}, it has been frequently observed at a variety of wavelengths. This object has a central source enshrouded by thick ejecta, which, on one hand, precludes a clear view of the central engine, but, on the other hand, displays dynamical and chemical indications of a probable hypernova progenitor \citep{Paczynski:1998pL45,Smith:2007p1116}.

The spectrum of $\eta$~Car is rich in low excitation forbidden and permitted lines \citep{Thackeray:1953p211}, as well as high excitation forbidden ones \citep[][and references therein]{Damineli:1998p299}, which temporarily disappeared in 1948, and again in 1965, 1981, 1987 and 1992. These ``spectroscopic events'' \citep{Gaviola:1953p234, Rodgers:1967p99, Thackeray:1967p51, Zanella:1984p79} or ``low excitation events'' \citep{Damineli:1998p299} were originally believed to be part of S Doradus cycles, commonly seen in other luminous blue variable stars (LBV) similar to $\eta$~Car. This interpretation seemed to be supported by the observations of the \ion{He}{1}~$\lambda10830$ line, which reached minimum \citep{Damineli:1996pL49} close to the maximum of the near-infrared light curve \citep{Whitelock:1994p364}. However, the spectroscopic events were demonstrated to be strictly periodic \citep{Damineli:2000pL101}, in contrast to the incoherent variations characteristic of  S Doradus oscillations.
\begin{figure*}[t]
\centering
\epsscale{1.13}
\plottwo{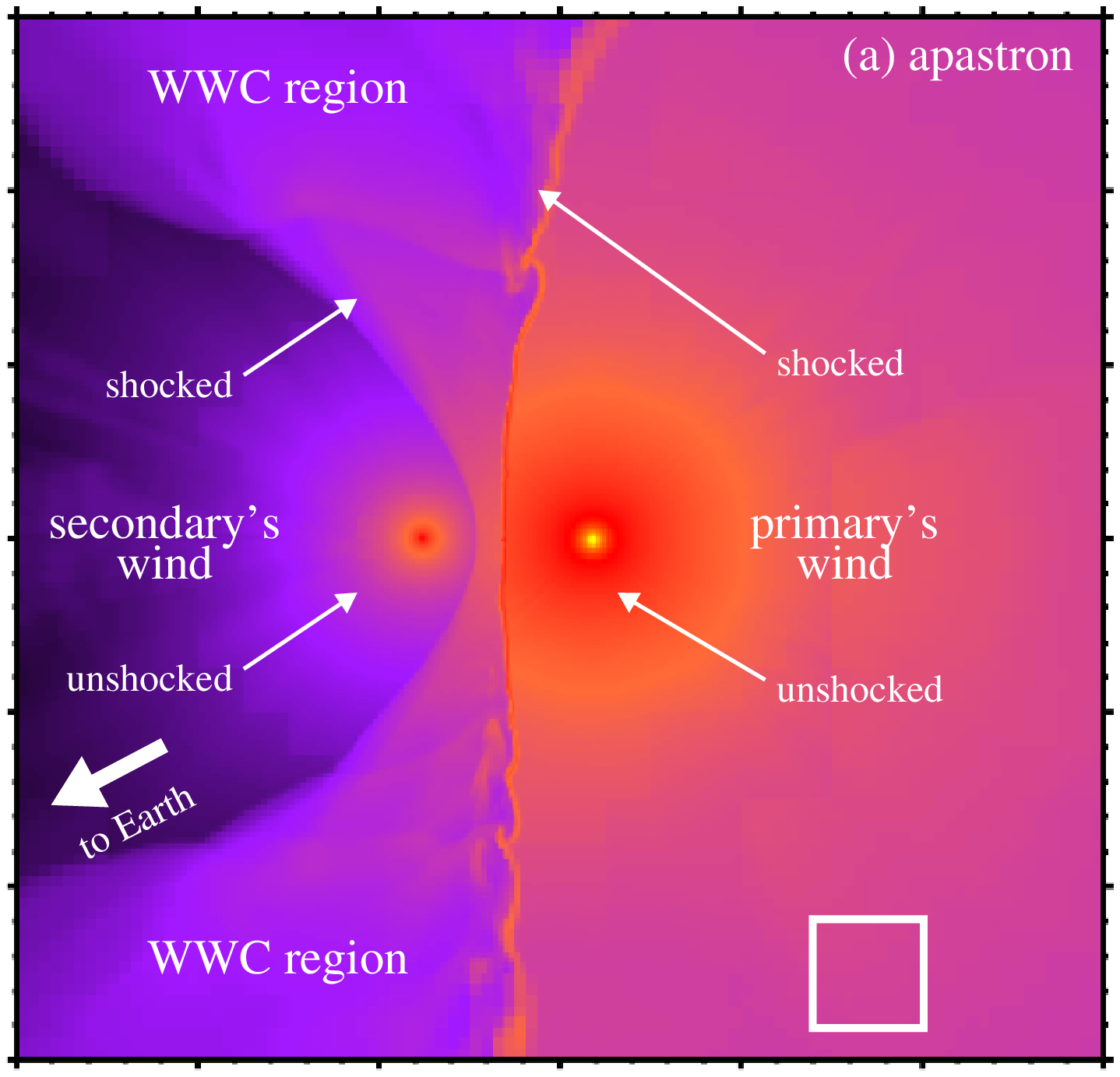}{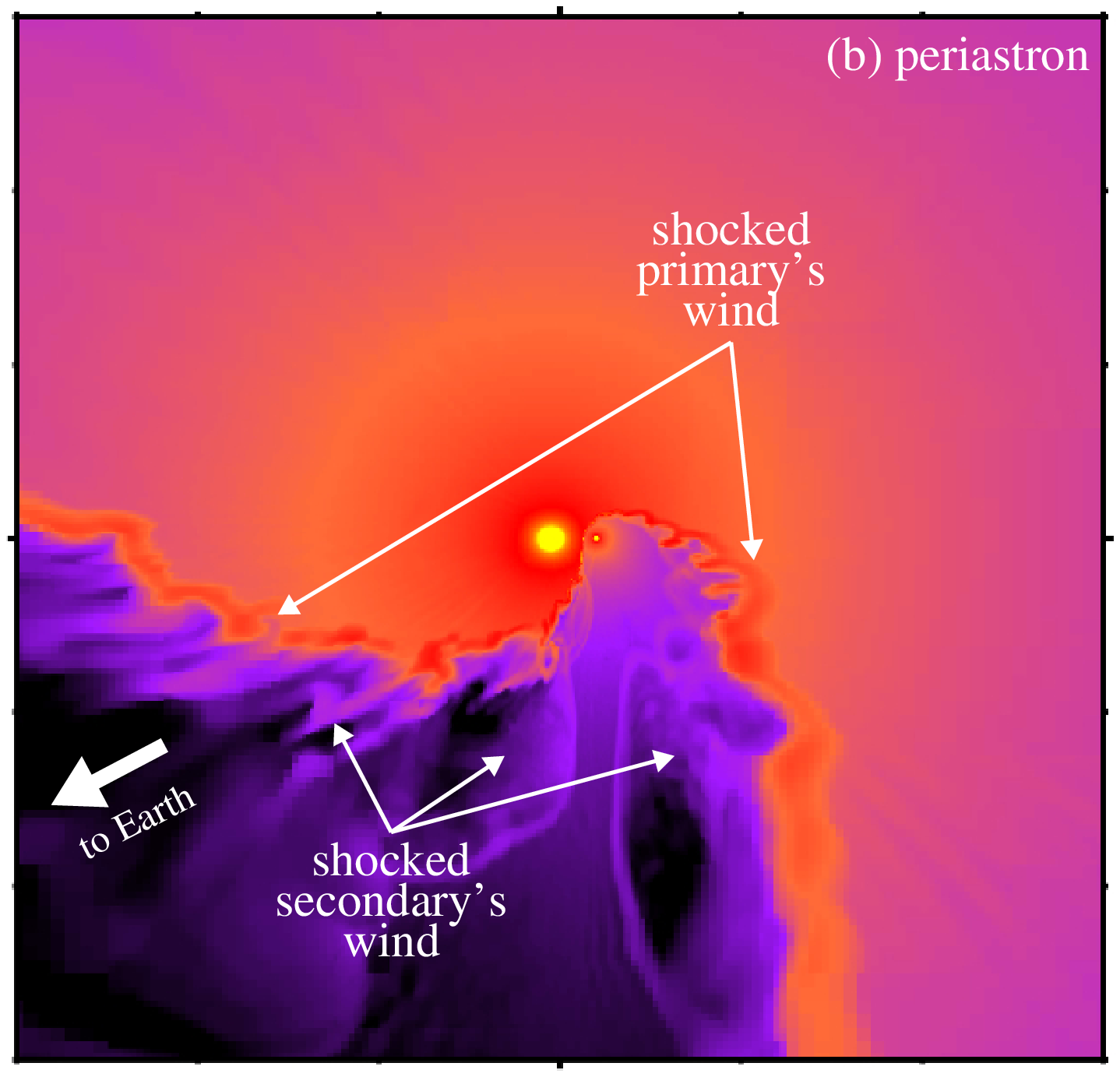}
\figcaption{\label{sketch}Snapshots of 3D simulations from \citet[][]{Parkin:2011p105}: (a) the inner $\pm15\times10^{14}$~cm of the system at apastron and (b) the inner $\pm3\times10^{14}$~cm of the system at periastron. For comparison, the small box at the bottom of panel (a) has the same dimensions as the entire figure in panel (b). In both panels, the thick arrow indicates the observer's line-of-sight corresponding to a longitude of periastron $\omega=243^\circ$ (the observer is located to the lower left).}
\end{figure*}

Periodic minima are in fact observed throughout the electromagnetic spectrum -- in doubly ionized forbidden, permitted and UV pumped lines in the optical and UV bands; in near-infrared continuum flux; in X-ray flux; and in mm and cm radio flux. The presence of a binary system \citep[][]{Damineli:1997p107} accounts for the strict periodicity of these minima \citep{Corcoran:2005p2018,FernandezLajus:2010p108,Damineli:2008p1649}. In the binary scenario, the primary star is the most luminous member of the system. It is moderately evolved (in the LBV phase) and has a slow ($\sim 500$~km\,\,s$^{-1}$) and dense wind in which the low excitation lines are formed \citep{Hillier:2001p837, Hillier:2006p1098}. The companion star is not detected in the spectrum, being less luminous, less massive, hotter and with a fast ($\sim 3000$~km\,\,s$^{-1}$) wind \citep{Corcoran:2005p2018}. The orbit is very eccentric and may even be higher than $e = 0.9$.  At apastron, the secondary star would be placed on our side of the primary, since  the Weigelt blobs -- which are circumstellar material located $\sim0.3$~arcsec from the central source at our side of the system \citep{Weigelt:1986pL5,Weigelt:1995p11} -- display a high excitation state during most of the orbital cycle \citep{Davidson:1997p335, Dorland:2004p1052, Smith:2004p405}. The high excitation forbidden lines are absent from the spectrum for a relatively brief time ($\sim$ 6 months) when the system is close to periastron \citep{Damineli:2008p2330}. Moreover, X-ray observations indicate that the column density ($N_H$) is lower toward apastron \citep{Hamaguchi:2007p522}, indicating that the WWC shock cone is opened toward our direction at those phases. However, some authors have suggested different system orientations \citep{Abraham:2005p922,Soker:2005p540,Kashi:2008p1751,Falceta-Goncalves:2009p1245}. Nevertheless, the binary scenario has gained support from many analyses \citep{Ishibashi:1999p983,Pittard:2002p636,Okazaki:2008pL39,Parkin:2009p1758,Parkin:2011p105}.

The suggestion of a wind-wind collision \citep[hereafter WWC;][]{Damineli:1997p107,Damineli:1997p272} explains the excess X-ray emission as compared to other binaries containing hot luminous components, especially at energies greater than 2 keV. \citet{Ishibashi:1999p983}, using the analytical model of \citet{Usov:1992p635}, were able to account for the main features in the X-ray emission and for the general behavior of the X-ray light curve, in particular, the increase in the X-ray flux just before periastron. They attributed the rapid drop in X-ray flux at periastron to an increase in the column density ($N_H$) of the intervening gas from the primary star in the line-of-sight, as the secondary star moved behind the primary. The duration of the X-ray minimum, however, was too long compared to the predictions of the analytical model, and an \emph{ad hoc} disk around the equator of the primary star was suggested as an additional source of obscuration \citep{Ishibashi:1999p983}. However, \citet{Damineli:1999p288} noted that an eclipse could not produce simultaneously a minimum in X-rays and in the high excitation lines, which are located at different distances and directions relative to our line-of-sight, but which disappear at nearly the same time.

Alternative models assumed transient phenomena around periastron to explain the disappearance of X-rays -- e.g., shell ejection \citep{Davidson:2005p900}, accretion onto the secondary companion \citep{Soker:2003p513,Soker:2005p540,Soker:2006p1563,Soker:2007p482}, but all of them face problems when confronted with the whole set of observational data. This is due to the fact that the spectroscopic cycle (i.e. the time interval between two consecutive disappearances of doubly ionized lines or narrow components in \ion{He}{1} lines) has two components: a ``slow variation" and a ``collapse" \citep{Damineli:2008p2330}. Each component affects the cycle in a different way. The ``slow variation" component is seen along the whole spectroscopic cycle as a smooth periodic modulation in the radio emission and in the intensity of low-excitation spectral features (e.g. singly ionized lines and the near infrared emission). The minimum of the ``slow variation" component is centered at $\phi\sim$ 0.07\footnote{In this paper,  phase $\phi$ is given by $\phi = {\rm (JD}-\rm{JD_0})/2022.7$, where JD$_0=2,452,819.2$ is the time when the narrow component of \ion{He}{1}~$\lambda6678$ reached its minimum value during event \#11.}. The ``collapse" component regulates the intensity of the high-energy features (e.g. X-rays, doubly ionized lines) with a minimum centered at $\phi\sim$ 0.03. Since the spectroscopic cycle is composed of these two components, any short-lived mechanism (shell ejection, accretion etc.) invoked to explain the ``collapse" would not be able to also explain the ``slow variation" component.

Modeling the wind-wind collision seems to be a viable option to derive the orbital parameters. Many models to explain the physics of the WWC have been presented, including 2D and 3D numerical simulations \citep{Pittard:2002p636,Okazaki:2008pL39,Parkin:2009p1758,Parkin:2011p105}. For reference, Fig.~\ref{sketch} shows the WWC from a 3D model \citet{Parkin:2011p105} in which we have identified the main regions and structures discussed in the following sections. \citet{Groh:2010p9} showed that a 3D numerical smoothed-particle hydrodynamics model \citep{Okazaki:2008pL39} could successfully explain the high velocity absorption component ($v\gtrsim2000$~km\,\,s$^{-1}$) seen in \ion{He}{1}~$\lambda10830$  near periastron. This high-velocity feature requires a large column density of high velocity gas  in the line-of-sight, and they derived a periastron longitude $\omega=240^\circ-270^\circ$ (consistent with an orientation such that the companion is on the near side of the primary at apastron) and orbital inclination $i=40^{\circ}-60^{\circ}$. The \ion{He}{1}~$\lambda10830$  line is emitted close to the WWC apex, which revolves very rapidly at periastron. Modeling of the spatially resolved forbidden line emission led to a similar longitude of periastron \citep{Gull:2009p1308, Gull:2011, Madura:2011}.

In $\eta$~Car, the detection of \ion{He}{2}~$\lambda4686$ emission \citep{Steiner:2004pL133} provided a new tool with which to study the wind-wind collision \citep{Martin:2006p474,Soker:2006p1563,Abraham:2007p309}. \citet{Steiner:2004pL133} found that this emission is present throughout the entire 5.538-yr cycle, but at a low intensity level (equivalent width, $|$EW$|<0.1$~\AA) for most of the orbit. The intensity of the line rises suddenly by one order of magnitude near $\phi=0$ and then sharply decreases to zero, after which the intensity reaches a lower peak and then declines. The intrinsic luminosity emitted in this spectral line (after correction for interstellar and circumstellar extinction) is greater than the maximum unabsorbed X-ray luminosity in the $2-10$ keV band \citep[$\sim67$~$L_\odot$,][]{Ishibashi:1999p983}. An ultraviolet source with $\sim10^4$~$L_\odot$ at energies higher than the He$^+$ ionization limit (54.4~eV) is needed to populate the upper level of this line. Keeping in mind that the line formation process for this line has a low efficiency \citep[$\sim$0.3 per cent, as derived by][and the present paper]{Martin:2006p474}, then the \ion{He}{2}~$\lambda4686$ line reveals a powerful transient energy source in the central region of the system, a peculiar state which encouraged us to study this spectral line in greater temporal detail.

The \ion{He}{2}~$\lambda4686$ line has been studied in other colliding wind binaries such as HD5980 \citep{Koenigsberger:2010p2600,Breysacher:2000p231,Moffat:1998p896}, V444~Cygni \citep{Marchenko:1997p826,Flores:2001p341}, WR127 \citep{delaChevrotiere:2011p635}, and R145 \citep{Schnurr:2009p823}. In those systems, it is not easy to isolate the contribution from the WWC, since the stellar winds themselves make an important contribution to the emission \citep[e.g. in the case of V444 Cygni, the WWC region contributes only $\sim10$~per cent of the \ion{He}{2}~$\lambda4686$ emission;][]{Marchenko:1997p826,Flores:2001p341}. Since the line emission is very faint far from periastron in $\eta$~Car, the emission around periastron can be associated entirely with the WWC.  There are other examples of spectral lines being formed exclusively in the WWC in other WC+O binaries, like the \ion{C}{3}~$\lambda5696$ excess in WR140 \citep{Fahed:2011p668}, Br22 \citep{Bartzakos:2001p33}, and $\theta$~Mus \citep{Hill:2002p1069}.

There are only two plausible regions in the system that could produce the observed \heii~emission near periastron in \ec: the shocked stellar wind of the primary or that of the secondary star. The latter seems insufficient to produce enough He$^{++}$ ions for recombination due to its high temperature and low density \citep[][]{Martin:2006p474} whereas the former seems to have the most favorable physical conditions (e.g. density, temperature and velocity field) for that process to occur.

\citet{Martin:2006p474} showed that UV photons with $h\nu>13.6$~eV could ionize He$^+$ excited to the n=2 level to He$^{++}$ and that Hydrogen-Ly$\alpha$ could also excite He$^+$ ions from the $n=2$ to $n=4$ level, which would produce the \ion{He}{2}~$\lambda4686$ line via electronic cascade. This could happen along the WWC region, where the gas has a relatively low temperature ($T<10^5$~K). Therefore, the \ion{He}{2}~$\lambda4686$ emission would be somewhat connected to the hard X-ray source, but not spatially coincident with it, and the temporal behavior of both emissions should be similar.  Also, the secondary star evacuates a cavity in the primary's wind near periastron, which could provide an additional source of UV photons from the inner wind of the primary. However, this emission would probably deliver much less energy than is required to power the \ion{He}{2}~$\lambda4686$ line: $|$EW$|\sim0.1$~\AA~\citep{Martin:2006p474}.

The main goal of our observing campaign is to constrain the physical emission mechanism and to determine the particular location of the \ion{He}{2}~$\lambda4686$ emitting region.  The present paper is organized as follows. In \S\ref{obs}, the observations and data processing are presented. In \S\ref{method}, the adopted methodology for measuring the equivalent width and velocities of the \ion{He}{2}~$\lambda 4686$ line are shown. In \S\ref{results} we present the results and in \S\ref{discussion} the discussion. Finally, our main conclusions are shown in \S\ref{conclusions}.

\begin{figure*}[t]
\centering
\epsscale{1}
\plotone{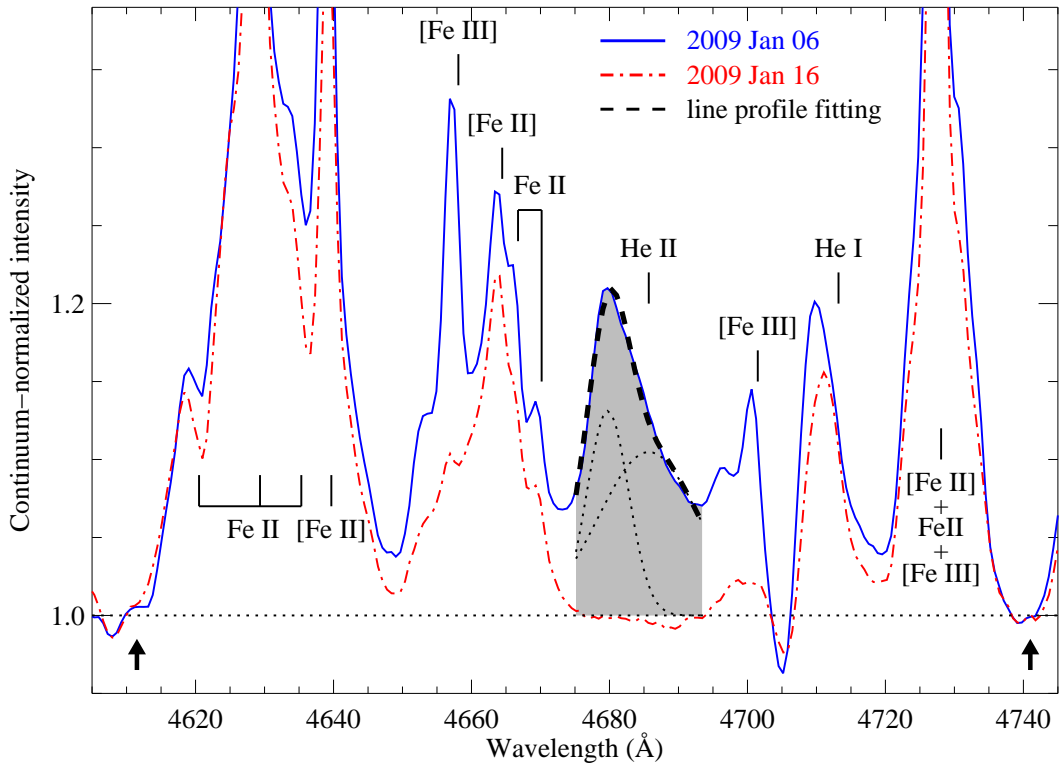}
\figcaption{\label{fig1}\ion{He}{2}~$\lambda4686$ at minimum (dash-dotted red line) and maximum (solid blue line). The observed line profile was fitted by two gaussian components (dotted lines) and the result (dashed line) was used to measure the radial velocity. The two arrows indicate the regions adopted as continuum level for the \ion{He}{2}~$\lambda4686$ emission line. The equivalent width was derived by direct integration under the line profile (gray shaded area).}
\end{figure*}

\section{Observations and data reduction}\label{obs}
We used the terminology of \citet{Groh:2004p1} in naming the cycles: \#1 starts in 1948 \citep[the first low excitation event recorded by][]{Gaviola:1953p234} and the last one in 2009, which is cycle \#12. In this paper, we use the term ``low excitation event" (or just ``event") to refer to the disappearance of high excitation spectral lines, which are used to define the spectroscopic cycle. We also used the term ``spectroscopic cycle" to refer to the time interval between two consecutive low excitation events \citep[5.538~yr;][]{Damineli:2008p2330}. The time of periastron passage is not known so we assign $\phi$ = 0 to the minimum of the narrow component of the \ion{He}{1}~$\lambda6678$ emission line, which occurred on JD=2,452,819,2 for event~\#11 \citep{Damineli:2008p1649}. We follow those authors also for the period length (P = 2022.7$\pm$ 1.3 days), determined by minimization of residuals when folding light curves around the minimum in cycles \#9, \#10 and \#11. This period is very robust, since it is based on the best sampled light curves, including the X-ray, optical, and near-infrared broad band light curves, and from variations in the strengths of several spectral lines. 

\subsection{Event \#12 (2009.0 minimum)}
The 2009.0 minimum was spectroscopically monitored using five astrophysical observatories: SOAR (Southern Astrophysical Research Telescope/Chile), OPD (Observat\'orio Pico dos Dias/Brazil), ESO (European Southern Observatory/Chile), CASLEO (Complejo Astron\'omico El Leoncito/Argentina), and LCO (Las Campanas Observatory/Chile). All the relevant data about the results presented in this paper are listed in Table~\ref{tbl-1}.

Daily monitoring of the optical spectrum of $\eta$~Car was carried out at SOAR from 2008 December 21 through 2009 January 23. After that, data were obtained once per month. The Goodman optical spectrograph was used to obtain spectra with spectral resolving power $R\sim2800$ (3~pixels) in the range $\lambda\lambda$3500--6850~\AA. All the observations were performed using the 0.45~arcsec-wide slit aligned with the parallactic angle. For each visit, a hot standard star \citep[HD~303308; \ion{O4.5}{5}((fc));][]{Walborn:2010pL143,Sota:2011p24} was observed to correct for telluric lines and for low-frequency variations in the detector response along the spectral dimension. Processing, reduction, and calibration of the SOAR data were done using the standard long-slit spectroscopic packages of {\sc iraf}\footnote{{\sc iraf} is distributed by the National Optical Astronomy Observatories, which are operated by the Association of Universities for Research in Astronomy, Inc., under cooperative agreement with the National Science Foundation.} \citep{Tody:1993p173}. The 1D spectrum was extracted using the {\sc iraf} package {\sc doslit} and the spatial extraction aperture was about 1.5~arcsec for all the SOAR spectra. Performing extractions with apertures as large as 4.5~arcsec does not introduce any significant change in either the \ion{He}{2}~$\lambda4686$ line profile or in the measured equivalent widths. Before or after each set of observations of $\eta$~Car, 2 exposures of an HgAr/CuAr lamp were obtained for wavelength calibration.

The FEROS (fiber-fed extended-range echelle spectrograph) data were obtained at the 2.2~m telescope (Max Planck Institute). FEROS delivers data with resolving power $R\sim48000$ (2 pixels) and spectral coverage from 3700 to 8600~\AA~in a single exposure. It uses two fibers with diameter of 2~arcsec separated by 2.9~arcmin. The processing, reduction, and calibration were performed using the FEROS standard reduction pipeline, which is described in \citet{Kaufer:1999p8}.

Spectra were also measured at the Coud\'e focus of the 1.60~m telescope of the Pico dos Dias Observatory (OPD/Brazil), with $R\sim10000$ (2 pixels). For each visit, a hot star ($\theta$~Car; \ion{B0.2}{5}; \citealt{Hubrig:2008p287,Naze:2008p801}) was also observed to correct the shape of the continuum for the variations in the instrument's response. The size of the extraction aperture for the OPD data was about 2~arcsec. A ThAr spectrum was obtained immediately preceding or following the \ec\ observations for wavelength calibration.

We also obtained spectra at the 6.5~m Magellan telescope located at Las Campanas Observatory, using the \'echelle spectrograph  MIKE, with $R\sim25000$ (2 pixels) in the blue region. MIKE's data processing was performed by cutting from the raw \'echelle images only the orders that contained the \ion{He}{2}~$\lambda4686$ line and its continuum region (orders from 75 to 78). Using this approach, we could process, reduce, and calibrate the data using typical {\sc iraf} \'echelle routines. For those data, we used the spectrum of a hot standard star (HD~303308, the same one used for the SOAR dataset) for a first rectification of the low-frequency variations along the spectral dimension.

Spectroscopic data were also obtained with the EBASIM and REOSC spectrographs at the CASLEO Observatory 2.15~m telescope. Both are \'echelle spectrographs with $R\sim42000$ and $R\sim12000$, respectively. The data reduction and processing were performed using standard IRAF tasks for \'echelle data.

The signal-to-noise ratio of the data from the 2009.0 campaign was always higher than 100 for all these observations, frequently reaching S/N$\sim$250.

\subsection{Previous events: \#9 (1992.5), \#10 (1998.0) and \#11 (2003.5)}
The data from previous events shown in the present paper were obtained at the Coud\'e focus of OPD and with the FLASH/HEROS spectrograph at the ESO 0.52~m telescope (events \#9 and \#11; we would like to remark that the ESO dataset for event \#9, used in the present work, was wrongly assigned to OPD in \citealt{Steiner:2004pL133}).  Spectra were also taken on 1997 November and December, and 1998 February and March, at the Coud\'e focus of the 1.9~m telescope of the Mount Stromlo and Siding Spring Observatory (MSSSO) with $R\sim60000$ in the range from 4346 to 4935~\AA~\citep{McGregor:1999p236a}.

\section{Measurement methodology for the \ion{H\MakeLowercase{e}}{2}~$\lambda4686$ line}\label{method}
\begin{figure*}[t]
\centering
\epsscale{1}
\plotone{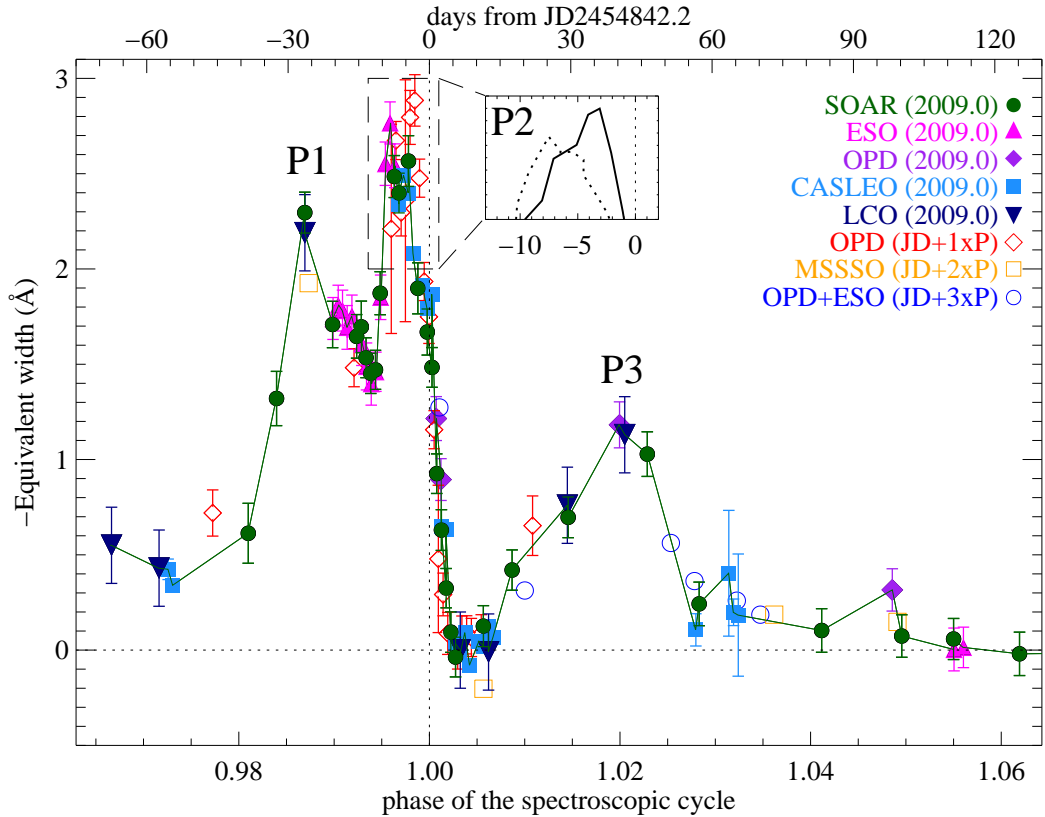}
\figcaption{\label{heii_ew}Equivalent width of \ion{He}{2}~$\lambda4686$ for the 2009.0 event (filled symbols) and for past events (open symbols) taken at different observatories. Data from the 3 previous events are over plotted as empty symbols, and folded with a period of P=2022.7~days. Each local peak discussed in the text are labeled in this figure as P1, P2 and P3. The inset panel shows the mean value of the equivalent width around peak P2 for event \#11 (solid line) and \#12 (dotted line) and illustrates that the peaks do not seem to be phase-locked. Indeed, for event \#12, P2 occurred $\sim4$~days before the phase observed for event \#11.}
\end{figure*}

In order to have a homogeneous dataset and to minimize systematic errors, we first identified the region to be considered as the \ion{He}{2}~$\lambda4686$ reference continuum. Fig.~\ref{fig1} illustrates the regions we adopted for the integration and continuum, which were set to a small wavelength interval (3~\AA~wide) near 4610~\AA~and 4740~\AA. Since the spectra were previously flattened using the spectrum of a hot star, the continuum in this range is well represented by a linear fit, and the equivalent width was measured by direct integration of the line profile in the range from 4675 to 4694~\AA. The regions defined in this work for the line integration and continuum are very similar to those adopted by \citet{Martin:2006p474} using \ \emph{Hubble Space Telescope} (\emph{HST}) data.

In order to have a robust determination of the radial velocity of the line peak, we used a two gaussian model to fit the integration region of the line profile, since a single gaussian profile did not reproduce the line profile over the integration region across the low excitation event, especially when the equivalent width reaches its maximum value near $\phi=0$. Fig.~\ref{fig1} shows the two gaussian components used to fit the \ion{He}{2}~$\lambda4686$ line profile.  The resulting 2-component fit was then used to find the wavelength at which the line profile reaches its peak intensity. Note that, since we are interested only in the radial velocity of the line peak, the line wings, which are much more extended, do not influence the derived velocity of the line peak. All of the radial velocities measured in the present paper are in the heliocentric frame.

The main source of uncertainties in the methodology for measuring the equivalent width is the normalization of the continuum. Data obtained in \'echelle mode suffer from order overlap and corrections for this overlap are crucial for tracking the real stellar continuum. We normalized the echelle spectra in two steps. First, we divided the flat-fielded spectrum by the continuum of a hot star, then we applied a parabolic fit to finally rectify the continuum. We rejected data in which the rectified continuum showed low-frequency variations greater than 1 per cent.

As mentioned previously, the wavelength calibration was obtained using comparison lamps. It is impossible to ascribe a single representative error to the entire observational dataset because each instrument has its own characteristics, reduction procedure and calibration methods. However, we checked the reliability of our procedure by comparing multiple observations taken on different telescopes in the same night. Discrepancies in velocity are smaller than $50$~km\,\,s$^{-1}$. Equivalent width measures from different telescopes at similar phases are also in quite good agreement.

We claim that our set of \ion{He}{2}~$\lambda4686$ spectra is the most accurate reported so far. It has the best time sampling 
over the last two events (\#11 = 2003.5 and \#12 = 2009.0) in addition to some data for cycles \#9 and \#10. Also, ground based spectra -- because they are spatially unresolved -- are very robust for time variation studies, since most of the emission comes from $\sim1$~arcsec around the central source, which is blurred by the seeing so that the ground based slit is always sampling the relevant emitting region. 

The absence of \ion{He}{2}~$\lambda4686$ emission on MJD~52852 reported by \citet{Martin:2006p474} might be caused by a bad datum or a real momentary fading of the line. Despite the fact that we cannot directly compare the results because we do not have simultaneous ground-based data, we call attention to the fact that our monitoring along 3 cycle (\#9, \#11 and \#12) always show this line in emission, confirming that P3 is a repeatable feature. Anyway, this point can be clarified in the next event (in 2014.6).

\section{Results}\label{results}
The observed \heii\ equivalent width curve is shown in Fig.~\ref{heii_ew}, which presents data for the last four events (\#9, \#10, \#11 and \#12) using the following ephemerides: 
\begin{equation}
\mbox{JD}=\mbox{JD}_0+2022.7(E - 11),
\end{equation}
where JD$_0=2,452,819.2$ \citep[time of minimum intensity of \ion{He}{1}~$\lambda6678$;][]{Damineli:2008p1649} and $E$ is the cycle+phase count. We focused on a time interval ranging from 2 months before up to 4~months after $\phi$=0. In this paper, we use the word ``day" as a reference to the number of days before (indicated by a minus sign) or after (plus sign) $\phi$=0 for each cycle.

\begin{figure}[t]
\centering
\includegraphics[width=8.6cm, bb=35 10 480 343, clip=true]{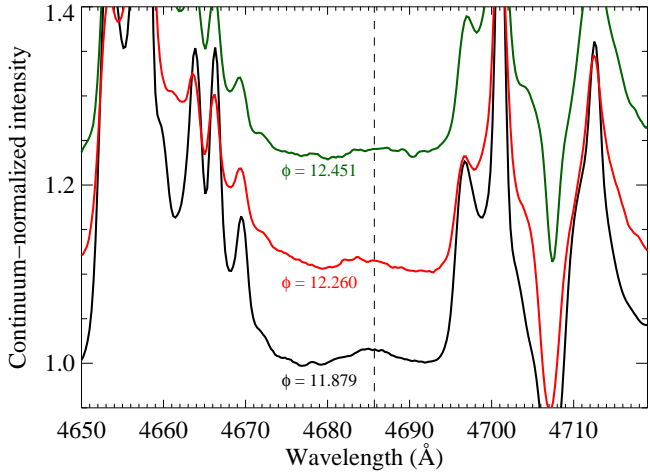}
\figcaption{\label{spectrumof2008May10}Spectra of $\eta$~Car taken on 2008 May 10 ($\phi=11.879$; black line), 2010 Jun 20 ($\phi=12.260$; red line), and 2011 July 11 ($\phi=12.451$), showing the presence of \ion{He}{2}~$\lambda4686$ at 8 months before, 17 months after and 31 months after the event, respectively (spectra taken at the OPD/Coud\'e). The corresponding phase of the spectroscopic cycle is also indicated. \ion{He}{2}~$\lambda4686$ far from periastron is present at faint level,  EW$< -0.1$~\AA.}
\end{figure}

\begin{figure}[t]
\centering
\includegraphics[width=0.5\textwidth, bb=-10 -10 535 360, clip=true]{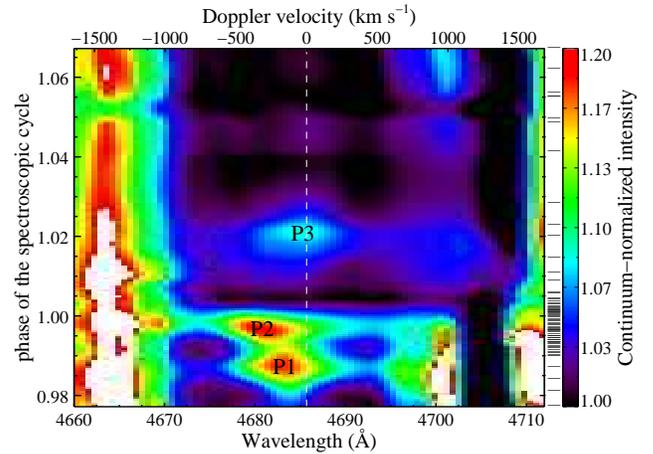}
\figcaption{\label{heii_stack}Folded time series spectra of the \ion{He}{2}~$\lambda4686$ line profile along the 4 last low excitation events. The continuum-normalized intensity of the spectra is color-coded between 0 and 20 per cent of the continuum level for visualization purposes. The vertical dashed line marks the rest wavelength of  the \ion{He}{2}~$\lambda4686$ line emission. The tick marks located on the right, between the color bar and the image, indicate the phases when spectra were obtained.}
\end{figure}

The 2003.5 spectra used to measure the equivalent width and radial velocities shown in this paper are the same as those presented in \citet{Steiner:2004pL133} but reduced according to the procedures described in \S \ref{obs} in order to keep the whole set of data as homogeneous as possible. Fig.~\ref{spectrumof2008May10} shows that \ion{He}{2}~$\lambda4686$ was detected with EW$\sim-0.1$\,\AA~in a spectrum taken at the OPD telescope on 2008 May 08, about 8 months before the low excitation event, and also at the same strength in another spectrum taken 17 months after $\phi$=0, on 2010 June 20 (using the same telescope and instrumental configuration). Another spectrum, taken on 2011 July 11, at $\phi = 12.451$ -- almost apastron -- also shows a very faint emission, which may be \ion{He}{2}~$\lambda4686$. These observations\footnote{Notice that none of them are included in Fig.~\ref{heii_ew} because they are far from the low excitation event.} suggest that \ion{He}{2}~$\lambda4686$ is present, except for a 5~day interval after the start of the spectroscopic event, along the whole 5.538-yr cycle \citep{Steiner:2004pL133}, though \citet{Martin:2006p474} claimed that this line is not present at any phase outside the low excitation event. The observed continuum-normalized time series spectrum of $\eta$~Car in the region $\lambda\lambda4660-4710$ is shown in Fig.~\ref{heii_stack}. In that figure, we interpolated the spectra in phase to show the temporal evolution of the \ion{He}{2}~$\lambda4686$ emission line. Notice that Fig.~\ref{heii_stack} mainly reflects the progression of the cycle \#11 to \#12 since data are sparse for other cycles.

A major feature in the intensity curve for event \#12 is the presence of 3 local maxima (labeled in Fig.~\ref{heii_ew} and \ref{heii_stack} as P1, P2 and P3). The first local maximum, P1, occurred about day $-26$, and showed a conspicuous peak with EW$=-2.30$~\AA. The second local maximum, P2, occurred between day $-10$ and $-5$, and during this time interval, the mean equivalent width was about $-2.5$~\AA. After day $-5$, the equivalent width of the line rapidly decreased to zero and it completely disappeared by day $+5$. From day $+5$ to $+10$ the equivalent width was zero. After day $+10$ it started to increase again, to a local maximum value of $-1.2$~\AA~(P3) at day $+40$, and then decreased to $-0.1$~\AA~after day $+115$. The rate of decrease of the EW just before minimum was about $-0.3$~\AA~per day and the $e$-folding time scale in the fading phase (after P2) was $\sim2$~days.

There are only a few sparse data near the low excitation event for cycles earlier than cycle \#12 (except for cycle \#11 in 2003.5, as shown in \citealt{Steiner:2004pL133}); despite this we can see that the general behavior of the equivalent width of the \ion{He}{2}~$\lambda4686$ line repeats faithfully from cycle to cycle. Peaks P1 and P2, however, do not seem to be phase-locked: for event \#12, P2 (the best time sampled peak) occurred $\sim4$ days before the phase in which it was detected in event \#11 (see inset in Fig.~\ref{heii_ew}). Although there were no data in the earlier cycles corresponding to the phase of the peak P3 maximum, the behavior of the line equivalent widths observed in the 1992.5 and 2003.5 events (see Fig.~\ref{heii_ew}) 
suggests that P3 is a phase-locked feature.

\subsection{\ion{He}{2}~$\lambda4686$ luminosity}
The intrinsic line luminosity $L^0$ of \ion{He}{2}~$\lambda4686$ is given by
\begin{equation}
L^0 = 4\pi D^2 f^0_{\lambda4686}\,{\rm EW_{He\,II\,\lambda4686}},
\label{eq:lint}
\end{equation}
where $D$ is the distance to the system (2.3~kpc), EW$_{He\,II\,\lambda4686}$ is the equivalent width of the \ion{He}{2}~$\lambda4686$ line (measured in the present work), and $f^0_{\lambda4686}$ is the intrinsic flux from the central source near $\lambda=4686$~\AA. 
We assume that $f^0_{\lambda4686}\approx f^{0}_{B}$, where $f^{0}_{B}$ is the dereddened $B$-band flux from the source.  The observed $B$-band flux ($f^{\rm obs}_{\rm B}$) is derived from observed $B$-band magnitudes following this relation:
\begin{equation}
f^{\rm obs}_{\rm B}=6.5\times10^{-10} \times 10^{-0.4B},
\end{equation}
where $B$ is the observed $B$-band brightness of the central source, as obtained by \citet{FernandezLajus:2009p1093}. It is important to remark that \citet{FernandezLajus:2009p1093} used a $12''$ radius aperture, which includes the entire Homunculus -- an important source of scattered light. In order to match the $1''$~region sampled by the ground based spectroscopy, we used a $B$~band image to determine the magnitude difference between a $1''$~circular aperture centered on the star and the $12''$~aperture used for the rest of the photometry. This showed that the $B$~band magnitudes reported by \citet{FernandezLajus:2009p1093} needed to be fainter by $+1.6$~mag to match the spectral aperture. The dereddened flux at 4686~\AA\ is then given by 
\begin{equation}
f^0_{\lambda4686}\approx f^0_B = f^{\rm obs}_B \times 10^{0.4A_B},
\end{equation}
where $A_B$ is the extinction in the $B$ band. However, since we have no direct measurement of $A_B$, we adopt the usual value for $A_V$ \citep{Davidson:1995p1784,Hillier:2001p837} to obtain $A_B$ using the following linear relation from \citet{Cardelli:1989p245}:
\begin{equation}\label{cardelli}
\frac{A_B}{A_V} = a(\lambda)+\frac{b(\lambda)}{R_V},
\end{equation}
where $a(\lambda)$ and $b(\lambda)$ are constants which depends on the wavelength and, in the present work, need to be evaluated in the $B$ band region ($\lambda=4500$~\AA).

The secular brightening of $\eta$ Car has been interpreted as a decrease in extinction \citep[see][]{vanGenderen:2006p3}, probably due to a expansion in the circumstellar shell, so we need to estimate the decrease of $A_V$ with time.  In 1998, in the line of sight to the central source there were about $A_V=7$ magnitudes of total extinction \citep[interstellar and circumstellar;][]{Hillier:2001p837} from which $\sim2$~magnitudes were due to gray extinction associated with large grains around the central source. Based on this result, in early 2006 the estimated extinction was $A_V$ = 5.7 \citep{vanGenderen:2006p3}. From 2003 to 2009, the average rate of increase in the brightness of the central source was $-0.06$~mag\,\,yr$^{-1}$. Therefore, the total extinction to the central source in 2003.5 was $A_V=5.9$, and in 2009.0, $A_V=5.5$. We do not correct for the scattering 
in the primary's wind, which should contribute to a cyclic variability in  $A_V$. 

In order to transform from $A_V$ to $A_B$, we used eq. (\ref{cardelli}), with $R_V=4.9$ \citep{Hillier:2001p837,vanGenderen:2006p3}. Hence, with the photometry in the $B$ band and the total extinction $A_B$, we were able to calculate the intrinsic flux $f^0_{\lambda4686}$ and thus the intrinsic luminosity of the \ion{He}{2}~$\lambda4686$ emission line from eq. (\ref{eq:lint}). 

The results are shown in Fig.~\ref{trio}a, where the intrinsic line luminosities are shown -- as well as the photon fluxes -- for two events: 2003.5 and 2009.0. The line luminosity of the 2003.5 event was higher than that of 2009.0 (for a given phase): the peak values for the intrinsic line luminosity were, respectively, 310~$L_\odot$ (corresponding to $1.8\times10^{47}$ \ion{He}{2}~$\lambda4686$ photons per second) and 250~$L_\odot$ (=$1.4\times10^{47}$ \ion{He}{2}~$\lambda4686$ photons per second). We cannot claim that there is a real difference in the intrinsic luminosity, since the extinction might have varied in shorter time scales than we adopted. These luminosities are probably lower limits to the true line luminosity since the adopted correction for extinction is only a simple estimate, and because obscuration by the wind of the primary star could also be significant.

\subsection{Radial velocities}

\begin{figure*}[t]
\centering
\epsscale{0.8}
\plotone{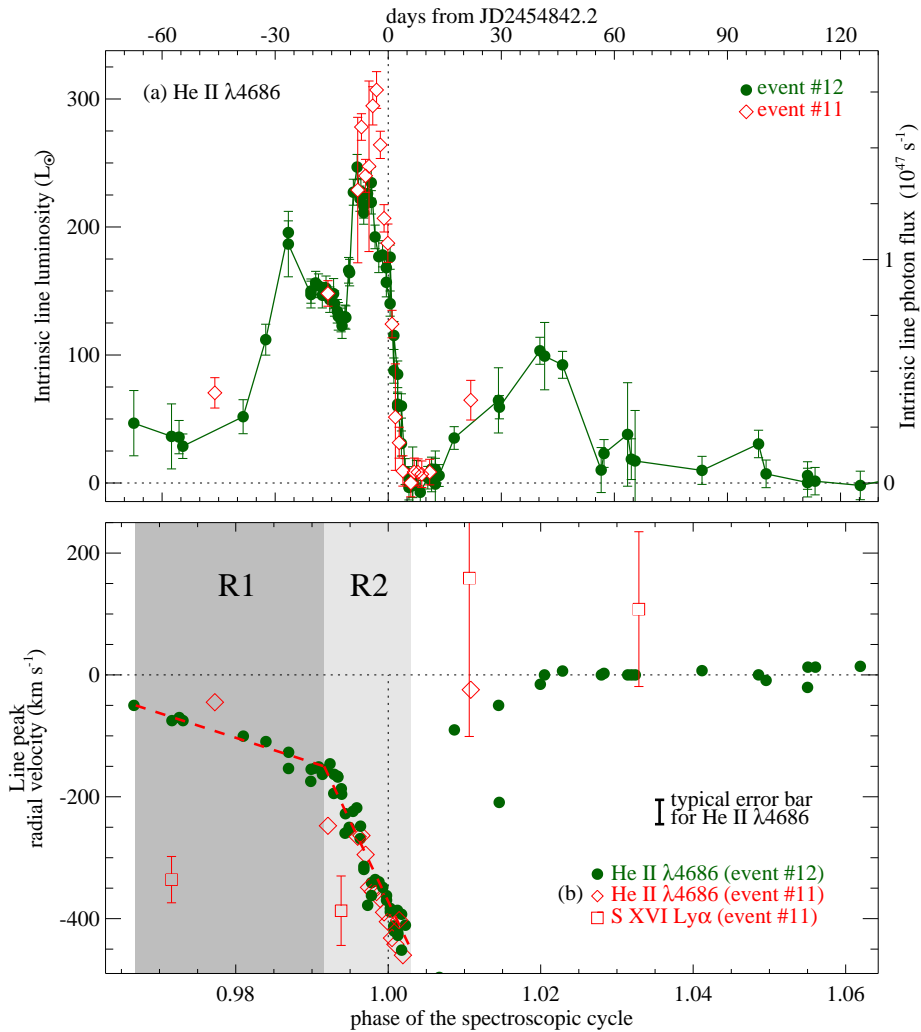}
\figcaption{(a) Lower limit for the intrinsic luminosity of the \ion{He}{2}~$\lambda$4686 line for 2009.0 (filled green circles) superimposed on the 2003.5 event (empty red diamonds) folded with a period of P=2022.7~days. (b) Radial velocity of the peak of the \ion{He}{2}~$\lambda4686$ line for both events and the observed velocity of the \ion{S}{16}~Ly$\alpha$ for the 2003.5 event \citep[red squares; data extracted from][]{Henley:2008p705}. The shaded areas indicate the two different regimes (R1 and R2) for the rate of change in the radial velocity curve. The abscissa in panel (a) is plotted in the same time scale as in (b).\label{trio}}
\end{figure*}

During most of the spectroscopic cycle, the radial velocity of the line peak was close to zero. Near the low excitation event, there were two regimes for the rate of change in the radial velocity (shaded areas R1 and R2 in Figure~\ref{trio}b). Region R1 is near day $-90$ when the radial velocity of the line peak changed at a daily rate of about $-2$~km\,\,s$^{-1}$ (shifting from $-50$~km\,\,s$^{-1}$ on day $-67$ to about $-150$~km\,\,s$^{-1}$ on day $-17$). The first peak in the equivalent width (P1) occurred during this time interval, on day $-26$. The second regime (R2) started on day $-17$ and was observed until the line completely disappeared, on day $+6$. During this time interval, the rate of change in the radial velocity of the line peak was almost seven times greater than in the previous interval: it started at $-150$~km\,\,s$^{-1}$ on day $-17$, and reached $-450$~km~s$^{-1}$ on day $+5$, a change of nearly 14~km\,\,s$^{-1}$ per day. Between day $+6$ and $+15$ the line could not be measured.

The general behavior of the \ion{He}{2}~$\lambda4686$ radial velocity suggests that this line is formed in the WWC region, in the interval from 2 months before to 2 months after the minimum. This is supported by the observed velocity of the X-ray line \ion{S}{16}~Ly$\alpha$ \citep[Figure~\ref{trio}b;][]{Henley:2008p705}. The time variations in radial velocity for both lines are similar: they become more blue-shifted near $\phi$ = 0 and then shift to projected velocities near zero. However, the velocity of the \ion{S}{16}~Ly$\alpha$ line just prior to the low-excitation event is more than a factor of 2 larger than that of \heii\ line. This is surprising, since we expect the \ion{S}{16}~Ly$\alpha$ line to be emitted close to the apex, where the flow velocity is slower than it is downstream along the WWC region. On the other hand, near $\phi$=0, the WWC gets very distorted and unstable, which could result in a mixed contribution from different regions.

The interpretation of the \ion{He}{2}~$\lambda4686$ radial velocity curve is not as simple as in other colliding wind systems because of the severe Coriolis distortion of the shock near periastron passage. As the stars approach periastron, the leading and the trailing arms of the shock are subject to different drag forces and the cone is heavily distorted. In addition, the inner zones, close to the apex, revolves fast in a spiraling pattern. Because of these complications, the projected radial velocities of the gas which produces the observable emission cannot be modeled using a simpler, rigid-cone geometry as has been used effectively for other colliding wind binary systems.

It is interesting to notice that the radial velocity of the peak of the \ion{He}{2}~$\lambda4686$ line emission shifts to the blue almost at the same time as the \ion{He}{1}~$\lambda10830$ P~Cyg absorption shows a high-velocity absorption component bluer than $-900$~km\,\,s$^{-1}$ \citep{Groh:2010p9}. Since the velocity field of the WWC region has a complex pattern close to periastron, we expect to see a wide range of velocities and this is in fact present in the large FWHM of those two lines. However, this interpretation may be too simple, since the absorption component samples a very limited region in our line of sight while the emission receives contributions from many other regions.

We should only see high positive velocities coming from the WWC region when it is pointing away from us, during periastron passage. Due to the high orbital eccentricity, the WWC cone does not rotate much until just before periastron passage, which means that the time spent by the WWC cone pointing away from the observer is very short.  From Fig.~\ref{trio}, it is evident that the maximum negative radial velocity occurs for P2, which is detected between day $-5$ and $-10$. Therefore, based on symmetry arguments for the motion of the secondary around the primary, the maximum positive radial velocity should occur somewhere between day +5 and +10, which corresponds to the interval when the \ion{He}{2}~$\lambda4686$ line is absent. Note, however, that this interval coincidently may occur at the same time as the disruption of the WWC region.

\begin{figure*}[t]
\centering
\epsscale{0.8}
\plotone{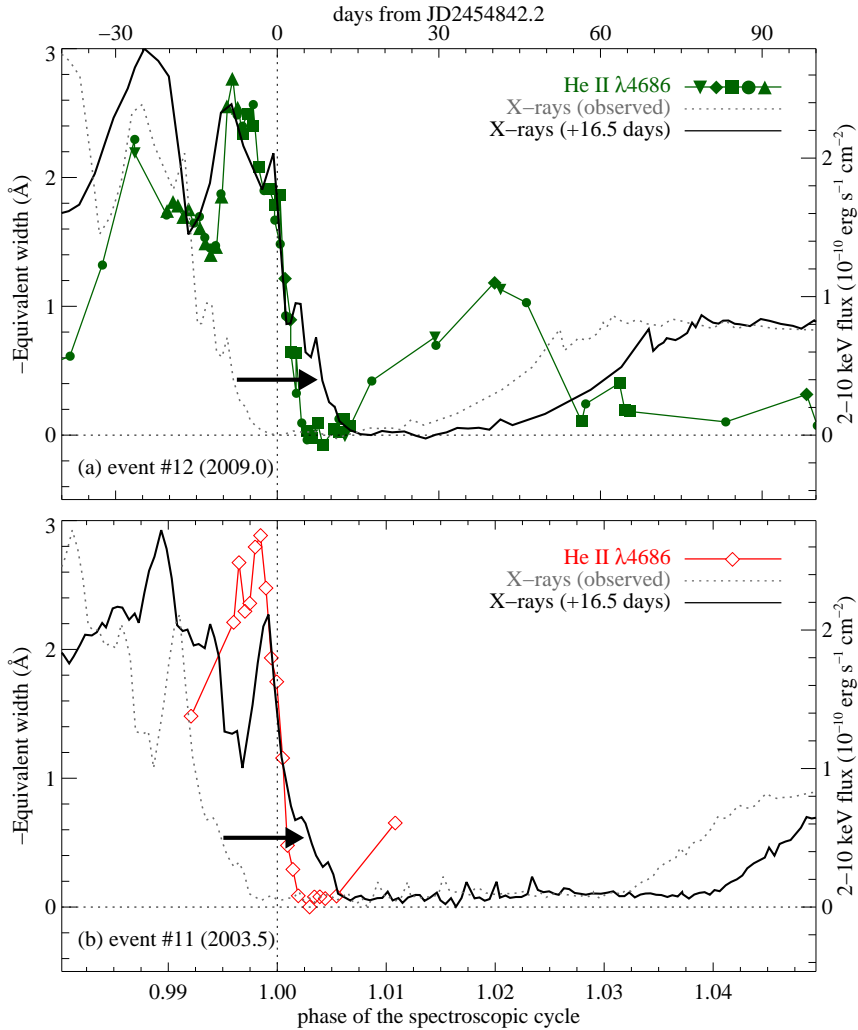}
\figcaption{\label{xrays}Equivalent width curve of \ion{He}{2}~$\lambda4686$ (symbols) over plotted with the 2--10~keV X-ray flux. In both panels, the observed X-ray flux is illustrated by the dotted gray line \citep[from][]{Corcoran:2005p2018}, while the solid black line was shifted by $+16.5$~days. Panel (a) is for cycle \#12 and (b) is for \#11.}
\end{figure*}

The radial velocity of the peak of the \ion{He}{2}~$\lambda4686$ line profile showed the same behavior for events \#11 and \#12: it reached up to $-450$~km\,\,s$^{-1}$ just before the minimum and completely vanished during 5 days (from day $+5$ to $+10$). After that, the peak was detected again with a radial velocity of $-100$~km\,\,s$^{-1}$, which changed to about zero $\sim$40 days after $\phi=0$.

\subsection{Correlation between \ion{He}{2}~$\lambda4686$ and X-rays}
During the 2009.0 low excitation event, after shifting the X-ray light curves in time by $+16.5$~days, the regime of fast decrease in the X-ray flux began approximately at the same time as the decrease in the \ion{He}{2}~$\lambda4686$ equivalent width  (see Fig.~\ref{xrays}a). The major peaks in the X-ray flux preceding the event also seem to have counterparts in the \ion{He}{2}~$\lambda4686$ equivalent width peaks. The 2003.5 observations did not show \emph{clearly} the same correlation between the X-ray flares and \heii\ peaks (see Figure~\ref{xrays}b), possibly because of sparser monitoring and/or lower quality \ion{He}{2}~$\lambda4686$ data. This suggests a possible connection between the X-ray and the \ion{He}{2}~$\lambda4686$ behavior. The delay of 16.5~days could be interpreted as the flow time of the shocked material, from the region where X-rays are created (in the WWC apex) to that where the temperature has dropped below $6\times10^5$~K, allowing for the recombination of He$^{++}$ into He$^{+}$. If so, assuming a flow velocity of $500$~km\,\,s$^{-1}$ for the shocked material \citep[terminal velocity of the primary's wind;][]{Hillier:2001p837,Pittard:2002p636}, then the bulk of the \ion{He}{2}~$\lambda4686$ emission should come from a region $4-5$~AU from the apex, which is consistent with the results of 3D models \citep{Parkin:2011p105}, which predict a temperature of $\sim10^5$~K in the primary's shocked stellar wind at that distance from the apex.

It is interesting to notice that the X-ray light curve of WR140\footnote{This system is the archetypal of the colliding wind binaries and is composed of a WC7+O4-5 binary in a highly eccentric ($\sim0.896$), long period ($\sim 2896.5$~days) orbit \citep[see][]{Williams:1990p662, White:1995p352, Marchenko:2003p1295, Fahed:2011p668}.} is similar to that of $\eta$~Car in the sense that there is a gradual increase until a maximum, followed by a sharp decrease \citep{Williams:2011p595, Corcoran:2011}. The main difference between these X-ray light curves is the duration of the X-ray minimum, which is much longer for $\eta$~Car (perhaps caused by the collapse of the WWC region, as discussed in the next section). Another important difference is the presence of X-rays flares in $\eta$ Car when approaching periastron, which are not seen in WR140; this might suggest that in the stellar wind of the WC7 primary component of WR 140, the clumps are smaller than those in the LBV primary star in \ec~\citep{Moffat:2009p693}.

\subsection{\ion{He}{2}~$\lambda4686$ far from periastron} \label{faint emission}
The faint emission seen along most of the orbit could be emitted in the WWC itself, since far from periastron, the density in the shock is smaller and the efficiency of \ion{He}{2}~$\lambda4686$ pumping is lower. However, since cooling is less efficient at lower density, the emission would occur farther from the apex, as compared to the strong emission at periastron. The speed of the shocked gas should be larger than it is near the apex, resulting in larger emission line velocities. Since at those phases the shock cone points in our direction, the observed radial velocity should be shifted to very negative values, different to what is observed, unless the opening angle of the cone increases by a large factor.

In principle, the \ion{He}{2}~$\lambda4686$ emission could be formed in the inner wind of the primary star, according to models computed by \citet{Hillier:2001p837, Hillier:2006p1098}. The problem then would be to explain its disappearance at periastron, since at that phase the primary star is at our side.

Another possibility is that the stellar wind of the secondary star would be the source of the faint \heii~ emission seen near apastron. Studies of the photoionization of the ejecta around the system suggest that the companion star has effective temperature between 37,000 and 43,000~K and luminosity between $10^5-10^6~L_\odot$ \citep{Mehner:2010p729, Verner:2005p973, Teodoro:2008p564}. As discussed by \citet{Hillier:2006p1098}, a companion star with $T_{\rm eff}=33,000$~K and $L=10^6~L_\odot$ would not have its spectral lines detected in the optical, since the flux of the primary star is at least a factor of $\sim 25$ higher than that of the companion at those wavelengths.

In any case, if the secondary star was the source of the faint \heii~ line observed far from periastron, the emission should  be stronger near apastron than far from it, since the WWC cavity is opened toward us at those phases \citep{Pittard:2002p636, Okazaki:2008pL39, Parkin:2009p1758, Parkin:2011p105}.

However, as seen in Fig.~\ref{spectrumof2008May10}, the emission line intensity at  $\phi = 0.45$ (close to apastron) is not stronger than at $\phi = 0.26$ or at $\phi = 0.897$, when the secondary is closer to periastron. Therefore, the only region left to be responsible for the production of the \heii~ at apastron is the WWC shock. In the next section, we will show that this line is formed in the WWC at periastron too.

\section{Discussion}\label{discussion}
Peak P2 (the absolute maximum) in the intrinsic luminosity of the \ion{He}{2}~$\lambda4686$ line emission corresponds to an emitted power of $1.2\times10^{36}$~erg\,\,s$^{-1}$ (recall that this is a lower limit because of uncertainties in the extinction correction). This large luminosity is emitted in only a single atomic transition. The secondary wind is already excluded as the source of the emission because the mechanism of illumination of its unshocked stellar wind by the hard UV/soft X-rays produced in the WWC region, proposed by \citet{Steiner:2004pL133}, does not account for the revised intrinsic photon flux of the \ion{He}{2}~$\lambda4686$ line. There are only two regions in which this enormous \heii\ luminosity could be produced: the shocked stellar wind of the primary or the shocked stellar wind of the secondary star.

Recent 3D models including inhibition effects \citep{Parkin:2009p1758} show that the wind of the secondary star close to periastron must collide with a slower pre-shock velocity ($\sim1,400$~km\,\,s$^{-1}$) due to the strong radiation field of the primary star (a process known as radiative inhibition). In this case, the shocked wind of the secondary near periastron would produce emission at lower energies than if the winds collided at terminal velocity. This softer emission might lead to an increase of flux near the peak of the He$^+$ ionization threshold at 0.0544~keV.

\begin{figure}[t]
\centering
\includegraphics[width=8.6cm, bb=35 15 480 343, clip=true]{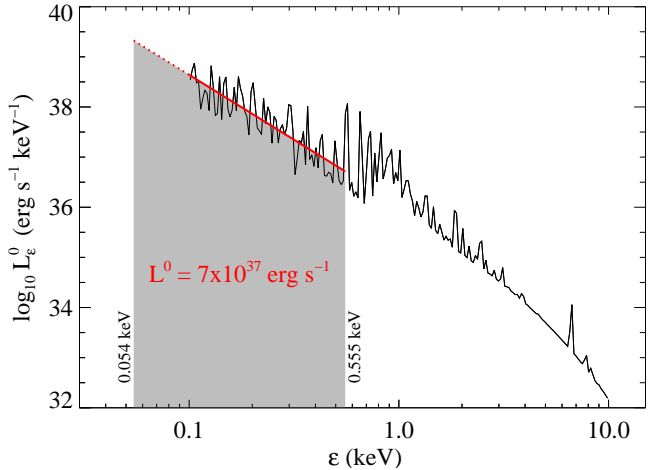}
\figcaption{\label{sed}Calculated spectral energy distribution (SED) of the shocked secondary's wind during periastron passage \citep[from][]{Parkin:2009p1758}. The intrinsic luminosity, $L^0$, was obtained by integrating the SED from 0.0544 to 0.550~keV (indicated by the gray shaded area). Notice that we fitted the SED (solid red line) and extrapolated it in the range from 0.0544 to 0.1~keV (dotted red line) because the output from simulations only goes down to 0.1~keV.}
\end{figure}

To test this, we adopted the results of 3D simulations \citep{Parkin:2009p1758,Parkin:2011p105} for the intrinsic spectral energy distribution (SED) of the shocked secondary's wind (Fig.~\ref{sed}) and used them to calculate the total power produced in the hard UV/soft X-ray domain during periastron passage (we extrapolated the SED down to 0.0544~keV and then integrated it from 0.0544 to 0.550~keV). This model predicts an intrinsic luminosity $L^0=7\times10^{37}$~erg\,\,s$^{-1}$ produced by the thermalization of the shocked secondary's wind. We then used the CLOUDY photoionization code to derive what fraction of the total \ion{He}{2} recombination emission would correspond to the transition from $n=4$ to $n=3$ in a high excitation nebula with density of $n_e\sim10^{10}$~cm$^{-3}$ (suitable for the shocked secondary's wind). We found that \ion{He}{2}~$\lambda4686$ accounts for less than 0.3~per cent of the total \ion{He}{2} emission. Therefore, the secondary's thermalized shocked wind would have to have an intrinsic luminosity of
\begin{equation}
L^0=\left(\frac{54.4}{4.2}\right) \left(\frac{L^{\rm obs}_{\lambda4686}}{0.003}\right)=5.2\times10^{39}~{\rm erg}\,\,{\rm s}^{-1}
\end{equation}
in order to account for the intrinsic luminosity of \ion{He}{2}~$\lambda4686$. This value is almost 2 orders of magnitude higher than that calculated by 3D models, and therefore, we take this as strong evidence that the \ion{He}{2}~$\lambda4686$ emission line does not originate in the secondary's shocked wind (even in the most favorable case where radiative inhibition causes the shock to occur at velocities lower than terminal).

The temperature, density and velocity of the primary's shocked stellar wind favor the production of photons with energy in the hard UV/soft X-ray domain suitable for ionizing or exciting He$^+$. Inside the high density ($n_e\sim10^{10}$~cm$^{-3}$) and low temperature ($<10^5$~K) shocked primary's wind, both the photoionization due to the X-ray and stellar radiation fields increase the extent of the He$^{++}$ region. This would naturally increase the number of recombining ions contributing to the line emission, which could account for the intrinsic luminosity of \ion{He}{2}~$\lambda4686$ near periastron. Furthermore, radiative effects in He$^+$ \citep{Martin:2006p474} can also increase the population in the $n=2$ level from where He$^{+}$ can be easily ionized by any photon with $h\nu>13.6$~eV (i.e. photons in the hydrogen Lyman continuum). Therefore, the intrinsic SED of the shocked primary's wind from 3D models would be of great interest to determine the exact amount of \ion{He}{2}~$\lambda4686$ emission within this region -- this study, however, is beyond the scope of the present paper. Therefore, based on the intrinsically high luminosity of the \ion{He}{2}~$\lambda4686$ emission line, we exclude all other possible regions but the primary's post-shocked wind as the region where the strong emission seen near periastron is formed.

The time delay of 16.5 days between the X-ray and the \ion{He}{2}~$\lambda4686$ light curves could be determined by the time flow of the gas from the WWC apex up to the location where  \ion{He}{2}~$\lambda4686$ is emitted. The maximum distance could be estimated by adopting the flow velocity as the primary's wind terminal speed ($500$~km\,\,s$^{-1}$), which amounts to $4-5$~AU from the apex. However, the real distance should be much smaller than that, since close to the apex the winds collide almost radially. The high density at periastron leads to very fast cooling, so that the \ion{He}{2}~$\lambda4686$ line should be formed not very far from the WWC apex. Using the \citet{Hillier:2001p837} model, we derive  impact parameter of the primary's wind in the broad band $B$-filter (which encompasses the \ion{He}{2}~$\lambda4686$ line)  as  D$_B=4.4$~AU.  Since the \ion{He}{2}~$\lambda4686$ emitting region is close to the WWC apex, which is very close to the primary star at periastron it could be eclipsed by the wind of the primary star for an intermediate orbital inclination. The expected duration of the eclipse of the \ion{He}{2}~$\lambda4686$ emitting region by the primary's stellar wind would be shorter than the X-ray emitting region, as the former is larger than the latter. In fact, the 2009.0 minimum in \ion{He}{2}~$\lambda4686$ lasted for a week as compared to the 4 weeks for the X-ray minimum. Both minima were centered at $\phi$=0.005, reinforcing the nature of the  \ion{He}{2}~$\lambda4686$ and the X-ray minima as due to occultation.

Soon after the minimum, the \ion{He}{2}~$\lambda4686$ starts rising again, reaching a local maximum (P3) in the phase range $\Delta$$\phi=0.01-0.03$ (a duration of 40 days). Interestingly, the $2-10$ keV X-ray emission increased very slowly at that time interval. We interpret this as a collapse of the WWC shock into deeper regions of the secondary stellar wind, where it is still being accelerated. As discussed by \citet{Parkin:2009p1758}, \citet{Corcoran:2010p1528} and \citet{Parkin:2011p105}, radiative inhibition of the secondary's wind might lead to a collapse of the WWC onto the ``surface" of the secondary when stars approach periastron. Of course, since the secondary star has a strong wind, it is likely that the momentum balance occurs at a radius larger than its photosphere.  Accretion onto the companion star might also play a role \citep{Soker:2006p1563,Soker:2007p482}, but is not a necessary condition to describe the observed phenomena. Thus, the secondary's wind is dominated by the radiation from the primary star (radiative inhibition) for a brief period and the hard X-ray source near the WWC apex is switched off. The associated  \ion{He}{2}~$\lambda4686$  emission may be absent during this time until the WWC forms anew. In the collapsed state, the WWC shock cone should be narrower, but still a powerful source at extreme UV energies,  since the stellar winds continue to collide downstream of the stagnation point.
 
\begin{figure}[t]
\centering
\includegraphics[width=8.6cm, bb=35 10 480 343, clip=true]{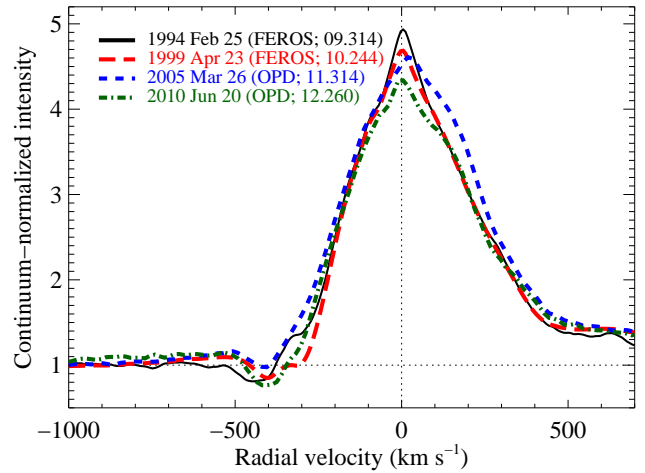}
\figcaption{\label{spec at same phase}Continuum-normalized line profile of H$\delta$ at approximately the same phase (indicated in the legend), along the last four cycles. Moderate variability at a low level is present, but no strong change  in the line strength (which would be expected if the mass loss rate dropped by a factor of 2) was seen in any of the four cycles.}
\end{figure}

As the stars separate after periastron passage, there should be a point in the orbit when the winds collide at their full terminal speeds once again. At that time, the WWC shock would be restored, and consequently the hard X-rays emission would restart. However, the wind of the secondary needs to do extra work to push back the primary's wind. This is probably not a smooth process, because the momentum equilibrium in the WWC apex does not only depend on the global properties of the winds, but also on local properties: clumps driven by radiative instabilities in the primary wind  \citep{Moffat:2009p693} might produce stochastic variability as the wind-wind shock front tries to reform. Changes in the clump properties could in principle explain why the minimum in X-rays was much shorter in cycle \#12 than in cycles \#11 and \#10, without needing to invoke a decrease by a factor of two in the mass-loss rate of the primary star \citep{Corcoran:2010p1528}. Also, lines formed deep in the primary's wind do not show particular changes in cycle \#12 compared to earlier cycles. Fig.~\ref{spec at same phase} displays line profiles of the line H$\delta$ taken almost at the same orbital phase ($\phi\sim0.3$) along the last 4 cycles (data obtained with FEROS/ESO and OPD/LNA). There are only small-scale variations in the P Cygni absorption strength and terminal speed. This line is relevant, since it is formed in regions of the primary's wind little affected by the WWC shock in the interval $\Delta$$\phi=0.2-0.8$, as can be seen in Fig.~\ref{orbit} \citep[which is based on the model by][]{Hillier:2001p837}. In this figure, the circles indicate the radii of maximum emissivity for different spectral lines. We also indicate the zone inside which the H$\delta$ emissivity is larger than 50~per cent of the peak intensity (dotted circle). H$\alpha$ is affected by the WWC for most of the orbit and \ion{Fe}{2} lines, formed at much larger distances from the central star\footnote{For example, the \ion{Fe}{2}~$\lambda4923$ emissivity is larger than 50 per cent of the peak emissivity within a ring which goes from $\sim25$ to $140$~A.U. from the primary star.}, are disturbed by the WWC even at apastron. In this way, changes reported by \citet{Mehner:2010pL22} in those lines cannot be attributed to the global properties of the primary's wind, but are probably driven by fluctuations in the WWC.

\section{Conclusions}\label{conclusions}
In this paper, we present data from the last 4 low excitation events in $\eta$~Car in the light of the \ion{He}{2}~$\lambda4686$ emission line. We summarize our results and conclusions below.

\begin{itemize}

\item The \ion{He}{2}~$\lambda4686$ line is present at a faint level along most of the 5.538 yr cycle;

\begin{figure}[t]
\centering
\includegraphics[width=8.6cm]{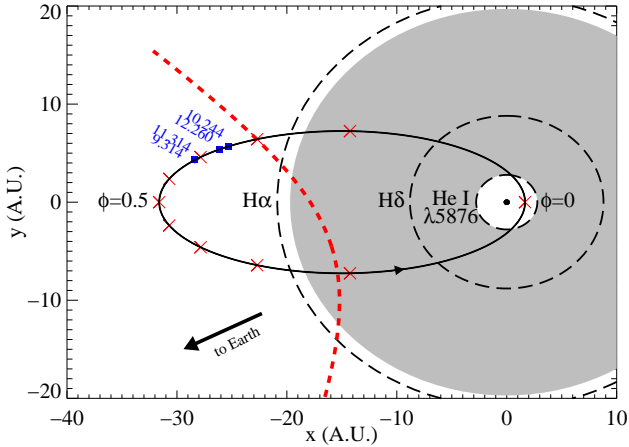}
\figcaption{\label{orbit}Projection of the orbital plane, relative to the position of the primary star ($e=0.9$ and $P=2022.7$~days). The concentric dashed circles correspond to the region of maximum emissivity of each labelled line. The gray shaded area indicates the zone where H$\delta$ emissivity is higher than 50 per cent of its maximum. The primary star is drawn to scale, with $R_*=60~R_\odot$. The red $\times$ symbols are drawn at each 0.1 in phase and the blue squares and numbers indicate the cycle+phase when the spectra shown in Fig.~\ref{spec at same phase} were obtained. The dashed paraboloid indicates the position of the WWC shock at $\phi$ = 0.8 and stellar mass of $90~M_\odot$ and $30~M_\odot$ for the primary and secondary star, respectively.}
\end{figure}

\item  two months before periastron, the \heii\ line strength increases dramatically with large flare-like oscillations on a timescale of weeks, similar to the X-ray ``flares" seen just prior to periastron passage;

\item  in the 2009.0 event, just before the minimum, there were 2 distinct peaks (P1 and P2) in the line equivalent width.  Peaks P1 and P2 are correlated with flares seen in the X-rays light curve, after shifting the later by $+16.5$~days. They are not phase-locked and are probably due to localized density enhancements (clumps) in the primary's stellar wind;

\item  there is an abrupt decrease in the line emission, reaching a minimum centered at $\phi=0.005$. This minimum is centered at the same phase as the X-ray minimum, but lasts only a week. We argue that this minimum is an occultation of the \ion{He}{2}~$\lambda4686$ emitting region by the primary's stellar wind;

\item  the fact that the disappearance of \ion{He}{2}~$\lambda4686$ emission in the central source and at the FOS4 position (in the SE lobe of the Homunculus) occurs close in time (but not simultaneously) is not an argument against an occultation, since the binary system revolves very quickly at periastron;

\item  in the 2009.0 event, after the minimum, the  \ion{He}{2}~$\lambda4686$ equivalent width shows another peak (P3), which appears to repeat from cycle to cycle. This peak is not correlated with the X-ray variation, as the \ion{He}{2}~$\lambda4686$ emission is in the high state when X-rays are very low. This \heii\ peak occurs in the phase range $\Delta\phi=1.01-1.035$;

\item  a potential mechanism to explain the P3 peak is a collapse of the WWC onto the ``surface" of the secondary star due to radiative inhibition of the secondary's wind by photons from the primary star, which could shift most of the energy of the shocked gas to lower energies near the He$^+$ ionization threshold;

\item  if the collapse scenario is valid, periastron occurs at $\phi\sim 0.02$ or a little earlier;

\item  the \ion{He}{2}~$\lambda4686$ equivalent width did not display the early recovery seen in X-rays after the 2009.0 event. We suggest that the recovery of the WWC region from the collapse phase is driven by localized instabilities/clumps in the primary's wind as suggested by \cite{Akashi:2006p451}. \citet{Moffat:2009p693} suggested that clumps might explain flares in the X-ray light-curve, and might control the momentum balance at scales on the clump size ($\sim1$ AU).

\end{itemize}

\acknowledgments
We thank the referee for questions and suggestions that led to a clarification of the ideas discussed in this paper. MT, AD and JES are grateful to the Brazilian agencies FAPESP and CNPq for continuous financial support. MT is supported through grants FAPESP 05/00190-8 and 09/08013-9. MBF acknowledges Conselho Nacional de Desenvolvimento Cient\'{\i}fico e Tecnol\'{o}gico (CNPq-Brazil) for the post-doctoral grant. MFC gratefully acknowledges support from NASA and Chandra via grants G07-8022A, G08-9018A, G09-0016A, and GO0-11039A, along with continued aid from the RXTE Guest Observer facility at NASA/GSFC. Calculations were performed with version 07.02 of Cloudy, last described by \citet{Ferland:1998p761}. This research has made use of NASA's Astrophysics Data System. In addition, this research has made use of data obtained from the High Energy Astrophysics Science Archive Research Center (HEASARC), provided by NASA's Goddard Space Flight Center.

{\it Facilities:} \facility{SOAR: 4.1m (Goodman), \facility{CASLEO: 2.15m (REOSC and EBASIM)}, \facility{ESO: 2.2m (FEROS)}, \facility{LNA: 1.6m (Coud\'e)}, \facility{Magellan: 6.5m Clay (MIKE)}, \facility{MtS: 1.9m (Coud\'e)} }.




\bibliography{formated}

\begin{thebibliography}{84}
\expandafter\ifx\csname natexlab\endcsname\relax\def\natexlab#1{#1}\fi

\bibitem[{Abraham \& Falceta-Gon{\c c}alves(2007)}]{Abraham:2007p309}
Abraham, Z. \& Falceta-Gon{\c c}alves, D. 2007, Monthly Notices of the Royal
  Astronomical Society, 378, 309

\bibitem[{Abraham {et~al.}(2005)Abraham, Falceta-Gon{\c c}alves, Dominici,
  Caproni, \& Jatenco-Pereira}]{Abraham:2005p922}
Abraham, Z., Falceta-Gon{\c c}alves, D., Dominici, T., Caproni, A., \&
  Jatenco-Pereira, V. 2005, Monthly Notices of the Royal Astronomical Society,
  364, 922

\bibitem[{Akashi {et~al.}(2006)Akashi, Soker, \& Behar}]{Akashi:2006p451}
Akashi, M., Soker, N., \& Behar, E. 2006, The Astrophysical Journal, 644, 451

\bibitem[{Bartzakos {et~al.}(2001)Bartzakos, Moffat, \&
  Niemela}]{Bartzakos:2001p33}
Bartzakos, P., Moffat, A. F.~J., \& Niemela, V.~S. 2001, Monthly Notices of the
  Royal Astronomical Society, 324, 33

\bibitem[{Breysacher \& Fran{\c c}ois(2000)}]{Breysacher:2000p231}
Breysacher, J. \& Fran{\c c}ois, P. 2000, Astronomy and Astrophysics, 361, 231

\bibitem[{Cardelli {et~al.}(1989)Cardelli, Clayton, \&
  Mathis}]{Cardelli:1989p245}
Cardelli, J.~A., Clayton, G.~C., \& Mathis, J.~S. 1989, ApJ, 345, 245

\bibitem[{Corcoran(2005)}]{Corcoran:2005p2018}
Corcoran, M.~F. 2005, The Astronomical Journal, 129, 2018

\bibitem[{Corcoran {et~al.}(2010)Corcoran, Hamaguchi, Pittard, Russell, Owocki,
  Parkin, \& Okazaki}]{Corcoran:2010p1528}
Corcoran, M.~F., Hamaguchi, K., Pittard, J.~M., Russell, C. M.~P., Owocki,
  S.~P., Parkin, E.~R., \& Okazaki, A. 2010, The Astrophysical Journal, 725,
  1528

\bibitem[{Corcoran {et~al.}(2011)Corcoran, Pollock, Hamaguchi, \&
  Russell}]{Corcoran:2011}
Corcoran, M.~F., Pollock, A. M.~T., Hamaguchi, K., \& Russell, C. 2011, eprint
  arXiv:1101.1422

\bibitem[{Damineli(1996)}]{Damineli:1996pL49}
Damineli, A. 1996, Astrophysical Journal Letters v.460, 460, L49

\bibitem[{Damineli(1997)}]{Damineli:1997p272}
---. 1997, Luminous Blue Variables: Massive Stars in Transition. ASP Conference
  Series; Vol. 120; 1997; ed. Antonella Nota and Henny Lamers (1997), 120, 272

\bibitem[{Damineli {et~al.}(1997)Damineli, Conti, \& Lopes}]{Damineli:1997p107}
Damineli, A., Conti, P.~S., \& Lopes, D.~F. 1997, New Astronomy, 2, 107

\bibitem[{Damineli {et~al.}(2008{\natexlab{a}})Damineli, Hillier, Corcoran,
  Stahl, Groh, Arias, Teodoro, Morrell, Gamen, Gonzalez, Leister, Levato,
  Levenhagen, Grosso, Colombo, \& Wallerstein}]{Damineli:2008p2330}
Damineli, A., Hillier, D.~J., Corcoran, M.~F., Stahl, O., Groh, J.~H., Arias,
  J., Teodoro, M., Morrell, N., Gamen, R., Gonzalez, F., Leister, N.~V.,
  Levato, H., Levenhagen, R.~S., Grosso, M., Colombo, J. F.~A., \& Wallerstein,
  G. 2008{\natexlab{a}}, Monthly Notices of the Royal Astronomical Society,
  386, 2330

\bibitem[{Damineli {et~al.}(2008{\natexlab{b}})Damineli, Hillier, Corcoran,
  Stahl, Levenhagen, Leister, Groh, Teodoro, Albacete~Colombo, Gonzalez, Arias,
  Levato, Grosso, Morrell, Gamen, Wallerstein, \& Niemela}]{Damineli:2008p1649}
Damineli, A., Hillier, D.~J., Corcoran, M.~F., Stahl, O., Levenhagen, R.~S.,
  Leister, N.~V., Groh, J.~H., Teodoro, M., Albacete~Colombo, J.~F., Gonzalez,
  F., Arias, J., Levato, H., Grosso, M., Morrell, N., Gamen, R., Wallerstein,
  G., \& Niemela, V. 2008{\natexlab{b}}, Monthly Notices of the Royal
  Astronomical Society, 384, 1649

\bibitem[{Damineli {et~al.}(2000)Damineli, Kaufer, Wolf, Stahl, Lopes, \&
  de~Ara{\'u}jo}]{Damineli:2000pL101}
Damineli, A., Kaufer, A., Wolf, B., Stahl, O., Lopes, D.~F., \& de~Ara{\'u}jo,
  F.~X. 2000, The Astrophysical Journal, 528, L101

\bibitem[{Damineli {et~al.}(1999)Damineli, Lopes, \& Conti}]{Damineli:1999p288}
Damineli, A., Lopes, D.~F., \& Conti, P.~S. 1999, Eta Carinae At The
  Millennium, 179, 288

\bibitem[{Damineli {et~al.}(1998)Damineli, Stahl, Kaufer, Wolf, Quast, \&
  Lopes}]{Damineli:1998p299}
Damineli, A., Stahl, O., Kaufer, A., Wolf, B., Quast, G., \& Lopes, D.~F. 1998,
  Astronomy and Astrophysics Supplement, 133, 299

\bibitem[{Davidson {et~al.}(1997)Davidson, Ebbets, Johansson, Morse, \&
  Hamann}]{Davidson:1997p335}
Davidson, K., Ebbets, D., Johansson, S., Morse, J.~A., \& Hamann, F.~W. 1997,
  Astronomical Journal v.113, 113, 335

\bibitem[{Davidson {et~al.}(1995)Davidson, Ebbets, Weigelt, Humphreys, Hajian,
  Walborn, \& Rosa}]{Davidson:1995p1784}
Davidson, K., Ebbets, D., Weigelt, G., Humphreys, R.~M., Hajian, A.~R.,
  Walborn, N.~R., \& Rosa, M. 1995, The Astronomical Journal, 109, 1784

\bibitem[{Davidson {et~al.}(2005)Davidson, Martin, Humphreys, Ishibashi, Gull,
  Stahl, Weis, Hillier, Damineli, Corcoran, \& Hamann}]{Davidson:2005p900}
Davidson, K., Martin, J., Humphreys, R.~M., Ishibashi, K., Gull, T.~R., Stahl,
  O., Weis, K., Hillier, D.~J., Damineli, A., Corcoran, M., \& Hamann, F. 2005,
  The Astronomical Journal, 129, 900

\bibitem[{de~la Chevroti{\`e}re {et~al.}(2011)de~la Chevroti{\`e}re, Moffat, \&
  Chen{\'e}}]{delaChevrotiere:2011p635}
de~la Chevroti{\`e}re, A., Moffat, A. F.~J., \& Chen{\'e}, A.~N. 2011, Monthly
  Notices of the Royal Astronomical Society, 411, 635

\bibitem[{Dorland {et~al.}(2004)Dorland, Currie, \& Hajian}]{Dorland:2004p1052}
Dorland, B.~N., Currie, D.~G., \& Hajian, A.~R. 2004, AJ, 127, 1052

\bibitem[{Fahed {et~al.}(2011)Fahed, Moffat, Zorec, Eversberg, Chen{\'e},
  Alves, Arnold, Bergmann, Gouveia~Carreira, Marques~Dias, Fernando,
  Sanchez~Gallego, Hunger, Knapen, Leadbeater, Morel, Rauw, Reinecke, Ribeiro,
  Romeo, dos Santos, Schanne, Stahl, Stober, Stober, Correia~Viegas, Vollmann,
  Corcoran, Dougherty, Pittard, Pollock, \& Williams}]{Fahed:2011p668}
Fahed, R., Moffat, A. F.~J., Zorec, J., Eversberg, T., Chen{\'e}, A.~N., Alves,
  F., Arnold, W., Bergmann, T., Gouveia~Carreira, L.~F., Marques~Dias, F.,
  Fernando, A., Sanchez~Gallego, J., Hunger, T., Knapen, J., Leadbeater, R.,
  Morel, T., Rauw, G., Reinecke, N., Ribeiro, J., Romeo, N., dos Santos, E.~M.,
  Schanne, L., Stahl, O., Stober, B., Stober, B., Correia~Viegas, N.~G.,
  Vollmann, K., Corcoran, M.~F., Dougherty, S.~M., Pittard, J.~M., Pollock, A.
  M.~T., \& Williams, P.~M. 2011, Soci{\'e}t{\'e} Royale des Sciences de
  Li{\`e}ge, 80, 668

\bibitem[{Falceta-Gon{\c c}alves \&
  Abraham(2009)}]{Falceta-Goncalves:2009p1245}
Falceta-Gon{\c c}alves, D. \& Abraham, Z. 2009, MNRAS, 1245

\bibitem[{Ferland {et~al.}(1998)Ferland, Korista, Verner, Ferguson, Kingdon, \&
  Verner}]{Ferland:1998p761}
Ferland, G.~J., Korista, K.~T., Verner, D.~A., Ferguson, J.~W., Kingdon, J.~B.,
  \& Verner, E.~M. 1998, PASP, 110, 761

\bibitem[{Fernandez~Lajus {et~al.}(2010)Fernandez~Lajus, Farina, Calderon,
  Salerno, Torres, Schwartz, von Essen, Giudici, \&
  Bareilles}]{FernandezLajus:2010p108}
Fernandez~Lajus, E., Farina, C., Calderon, J.~P., Salerno, N., Torres, A.~F.,
  Schwartz, M.~A., von Essen, C., Giudici, F., \& Bareilles, F.~A. 2010, New
  Astronomy, 15, 108

\bibitem[{Fernandez~Lajus {et~al.}(2009)Fernandez~Lajus, Farina, Torres,
  Schwartz, Salerno, Calderon, von Essen, Calcaferro, Giudici, Llinares, \&
  Niemela}]{FernandezLajus:2009p1093}
Fernandez~Lajus, E., Farina, C., Torres, A.~F., Schwartz, M.~A., Salerno, N.,
  Calderon, J.~P., von Essen, C., Calcaferro, L.~M., Giudici, F., Llinares, C.,
  \& Niemela, V. 2009, Astronomy and Astrophysics, 493, 1093

\bibitem[{Flores {et~al.}(2001)Flores, Auer, Koenigsberger, \&
  Cardona}]{Flores:2001p341}
Flores, A., Auer, L.~H., Koenigsberger, G., \& Cardona, O. 2001, The
  Astrophysical Journal, 563, 341

\bibitem[{Gaviola(1953)}]{Gaviola:1953p234}
Gaviola, E. 1953, ApJ, 118, 234

\bibitem[{Groh \& Damineli(2004)}]{Groh:2004p1}
Groh, J.~H. \& Damineli, A. 2004, Information Bulletin on Variable Stars, 5492,
  1

\bibitem[{Groh {et~al.}(2010)Groh, Nielsen, Damineli, Gull, Madura, Hillier,
  Teodoro, Driebe, Weigelt, Hartman, Kerber, Okazaki, Owocki, Millour,
  Murakawa, Kraus, Hofmann, \& Schertl}]{Groh:2010p9}
Groh, J.~H., Nielsen, K.~E., Damineli, A., Gull, T.~R., Madura, T.~I., Hillier,
  D.~J., Teodoro, M., Driebe, T., Weigelt, G., Hartman, H., Kerber, F.,
  Okazaki, A.~T., Owocki, S.~P., Millour, F., Murakawa, K., Kraus, S., Hofmann,
  K.-H., \& Schertl, D. 2010, Astronomy and Astrophysics, 517, 9

\bibitem[{Gull {et~al.}(2011)Gull, Madura, Groh, \& Corcoran}]{Gull:2011}
Gull, T.~R., Madura, T.~I., Groh, J.~H., \& Corcoran, M.~F. 2011, eprint
  arXiv:1110.6420

\bibitem[{Gull {et~al.}(2009)Gull, Nielsen, Corcoran, Madura, Owocki, Russell,
  Hillier, Hamaguchi, Kober, Weis, Stahl, \& Okazaki}]{Gull:2009p1308}
Gull, T.~R., Nielsen, K.~E., Corcoran, M.~F., Madura, T.~I., Owocki, S.~P.,
  Russell, C. M.~P., Hillier, D.~J., Hamaguchi, K., Kober, G.~V., Weis, K.,
  Stahl, O., \& Okazaki, A.~T. 2009, MNRAS, 396, 1308

\bibitem[{Hamaguchi {et~al.}(2007)Hamaguchi, Corcoran, Gull, Ishibashi,
  Pittard, Hillier, Damineli, Davidson, Nielsen, \& Kober}]{Hamaguchi:2007p522}
Hamaguchi, K., Corcoran, M.~F., Gull, T., Ishibashi, K., Pittard, J.~M.,
  Hillier, D.~J., Damineli, A., Davidson, K., Nielsen, K.~E., \& Kober, G.~V.
  2007, ApJ, 663, 522

\bibitem[{Henley {et~al.}(2008)Henley, Corcoran, Pittard, Stevens, Hamaguchi,
  \& Gull}]{Henley:2008p705}
Henley, D.~B., Corcoran, M.~F., Pittard, J.~M., Stevens, I.~R., Hamaguchi, K.,
  \& Gull, T.~R. 2008, The Astrophysical Journal, 680, 705

\bibitem[{Hill {et~al.}(2002)Hill, Moffat, \& St-Louis}]{Hill:2002p1069}
Hill, G.~M., Moffat, A. F.~J., \& St-Louis, N. 2002, Monthly Notices of the
  Royal Astronomical Society, 335, 1069

\bibitem[{Hillier {et~al.}(2001)Hillier, Davidson, Ishibashi, \&
  Gull}]{Hillier:2001p837}
Hillier, D.~J., Davidson, K., Ishibashi, K., \& Gull, T.~R. 2001, The
  Astrophysical Journal, 553, 837

\bibitem[{Hillier {et~al.}(2006)Hillier, Gull, Nielsen, Sonneborn, Iping,
  Smith, Corcoran, Damineli, Hamann, Martin, \& Weis}]{Hillier:2006p1098}
Hillier, D.~J., Gull, T., Nielsen, K., Sonneborn, G., Iping, R., Smith, N.,
  Corcoran, M., Damineli, A., Hamann, F.~W., Martin, J.~C., \& Weis, K. 2006,
  The Astrophysical Journal, 642, 1098

\bibitem[{Hubrig {et~al.}(2008)Hubrig, Briquet, Morel, Sch{\"o}ller,
  Gonz{\'a}lez, \& de~Cat}]{Hubrig:2008p287}
Hubrig, S., Briquet, M., Morel, T., Sch{\"o}ller, M., Gonz{\'a}lez, J.~F., \&
  de~Cat, P. 2008, Astronomy and Astrophysics, 488, 287

\bibitem[{Ishibashi {et~al.}(1999)Ishibashi, Corcoran, Davidson, Swank, Petre,
  Drake, Damineli, \& White}]{Ishibashi:1999p983}
Ishibashi, K., Corcoran, M.~F., Davidson, K., Swank, J.~H., Petre, R., Drake,
  S.~A., Damineli, A., \& White, S. 1999, The Astrophysical Journal, 524, 983

\bibitem[{Kashi \& Soker(2008)}]{Kashi:2008p1751}
Kashi, A. \& Soker, N. 2008, Monthly Notices of the Royal Astronomical Society,
  390, 1751

\bibitem[{Kaufer {et~al.}(1999)Kaufer, Stahl, Tubbesing, N{\o}rregaard, Avila,
  Francois, Pasquini, \& Pizzella}]{Kaufer:1999p8}
Kaufer, A., Stahl, O., Tubbesing, S., N{\o}rregaard, P., Avila, G., Francois,
  P., Pasquini, L., \& Pizzella, A. 1999, The Messenger, 95, 8

\bibitem[{Koenigsberger {et~al.}(2010)Koenigsberger, Georgiev, Hillier,
  Morrell, Barb{\'a}, \& Gamen}]{Koenigsberger:2010p2600}
Koenigsberger, G., Georgiev, L., Hillier, D.~J., Morrell, N., Barb{\'a}, R., \&
  Gamen, R. 2010, The Astronomical Journal, 139, 2600

\bibitem[{Madura {et~al.}(2011)Madura, Gull, Owocki, Groh, Okazaki, \&
  Russell}]{Madura:2011}
Madura, T.~I., Gull, T.~R., Owocki, S.~P., Groh, J.~H., Okazaki, A.~T., \&
  Russell, C. M.~P. 2011, eprint arXiv:1111.2226

\bibitem[{Marchenko {et~al.}(2003)Marchenko, Moffat, Ballereau, Chauville,
  Zorec, Hill, Annuk, Corral, Demers, Eenens, Panov, Seggewiss, Thomson, \&
  Villar~Sbaffi}]{Marchenko:2003p1295}
Marchenko, S.~V., Moffat, A. F.~J., Ballereau, D., Chauville, J., Zorec, J.,
  Hill, G.~M., Annuk, K., Corral, L.~J., Demers, H., Eenens, P. R.~J., Panov,
  K.~P., Seggewiss, W., Thomson, J.~R., \& Villar~Sbaffi, A. 2003, ApJ, 596,
  1295

\bibitem[{Marchenko {et~al.}(1997)Marchenko, Moffat, Eenens, Cardona,
  Echevarria, \& Hervieux}]{Marchenko:1997p826}
Marchenko, S.~V., Moffat, A. F.~J., Eenens, P. R.~J., Cardona, O., Echevarria,
  J., \& Hervieux, Y. 1997, Astrophysical Journal v.485, 485, 826

\bibitem[{Martin {et~al.}(2006)Martin, Davidson, Humphreys, Hillier, \&
  Ishibashi}]{Martin:2006p474}
Martin, J.~C., Davidson, K., Humphreys, R.~M., Hillier, D.~J., \& Ishibashi, K.
  2006, The Astrophysical Journal, 640, 474

\bibitem[{McGregor {et~al.}(1999)McGregor, Rathborne, \&
  Humphreys}]{McGregor:1999p236a}
McGregor, P.~J., Rathborne, J.~M., \& Humphreys, R.~M. 1999, Eta Carinae At The
  Millennium, 179, 236

\bibitem[{Mehner {et~al.}(2010{\natexlab{a}})Mehner, Davidson, Ferland, \&
  Humphreys}]{Mehner:2010p729}
Mehner, A., Davidson, K., Ferland, G.~J., \& Humphreys, R.~M.
  2010{\natexlab{a}}, The Astrophysical Journal, 710, 729

\bibitem[{Mehner {et~al.}(2010{\natexlab{b}})Mehner, Davidson, Humphreys,
  Martin, Ishibashi, Ferland, \& Walborn}]{Mehner:2010pL22}
Mehner, A., Davidson, K., Humphreys, R.~M., Martin, J.~C., Ishibashi, K.,
  Ferland, G.~J., \& Walborn, N.~R. 2010{\natexlab{b}}, The Astrophysical
  Journal Letters, 717, L22

\bibitem[{Moffat \& Corcoran(2009)}]{Moffat:2009p693}
Moffat, A. F.~J. \& Corcoran, M.~F. 2009, The Astrophysical Journal, 707, 693

\bibitem[{Moffat {et~al.}(1998)Moffat, Marchenko, Bartzakos, Niemela, Cerruti,
  Magalhaes, Balona, St-Louis, Seggewiss, \& Lamontagne}]{Moffat:1998p896}
Moffat, A. F.~J., Marchenko, S.~V., Bartzakos, P., Niemela, V.~S., Cerruti,
  M.~A., Magalhaes, A.~M., Balona, L., St-Louis, N., Seggewiss, W., \&
  Lamontagne, R. 1998, ApJ, 497, 896

\bibitem[{Naz{\'e} \& Rauw(2008)}]{Naze:2008p801}
Naz{\'e}, Y. \& Rauw, G. 2008, Astronomy and Astrophysics, 490, 801

\bibitem[{Okazaki {et~al.}(2008)Okazaki, Owocki, Russell, \&
  Corcoran}]{Okazaki:2008pL39}
Okazaki, A.~T., Owocki, S.~P., Russell, C. M.~P., \& Corcoran, M.~F. 2008,
  Monthly Notices of the Royal Astronomical Society: Letters, 388, L39

\bibitem[{Paczynski(1998)}]{Paczynski:1998pL45}
Paczynski, B. 1998, ApJS, 494, L45

\bibitem[{Parkin {et~al.}(2011)Parkin, Pittard, Corcoran, \&
  Hamaguchi}]{Parkin:2011p105}
Parkin, E.~R., Pittard, J.~M., Corcoran, M.~F., \& Hamaguchi, K. 2011, The
  Astrophysical Journal, 726, 105

\bibitem[{Parkin {et~al.}(2009)Parkin, Pittard, Corcoran, Hamaguchi, \&
  Stevens}]{Parkin:2009p1758}
Parkin, E.~R., Pittard, J.~M., Corcoran, M.~F., Hamaguchi, K., \& Stevens,
  I.~R. 2009, Monthly Notices of the Royal Astronomical Society, 394, 1758

\bibitem[{Pittard \& Corcoran(2002)}]{Pittard:2002p636}
Pittard, J.~M. \& Corcoran, M.~F. 2002, Astronomy and Astrophysics, 383, 636

\bibitem[{Rodgers \& Searle(1967)}]{Rodgers:1967p99}
Rodgers, A.~W. \& Searle, L. 1967, Monthly Notices of the Royal Astronomical
  Society, 135, 99

\bibitem[{Schnurr {et~al.}(2009)Schnurr, Moffat, Villar~Sbaffi, St-Louis, \&
  Morrell}]{Schnurr:2009p823}
Schnurr, O., Moffat, A. F.~J., Villar~Sbaffi, A., St-Louis, N., \& Morrell,
  N.~I. 2009, Monthly Notices of the Royal Astronomical Society, 395, 823

\bibitem[{Smith \& Frew(2011)}]{Smith:2011p2009}
Smith, N. \& Frew, D.~J. 2011, Monthly Notices of the Royal Astronomical
  Society, 415, 2009

\bibitem[{Smith {et~al.}(2007)Smith, Li, Foley, Wheeler, Pooley, Chornock,
  Filippenko, Silverman, Quimby, Bloom, \& Hansen}]{Smith:2007p1116}
Smith, N., Li, W., Foley, R.~J., Wheeler, J.~C., Pooley, D., Chornock, R.,
  Filippenko, A.~V., Silverman, J.~M., Quimby, R., Bloom, J.~S., \& Hansen, C.
  2007, The Astrophysical Journal, 666, 1116

\bibitem[{Smith {et~al.}(2004)Smith, Morse, Gull, Hillier, Gehrz, Walborn,
  Bautista, Collins, Corcoran, Damineli, Hamann, Hartman, Johansson, Stahl, \&
  Weis}]{Smith:2004p405}
Smith, N., Morse, J.~A., Gull, T.~R., Hillier, D.~J., Gehrz, R.~D., Walborn,
  N.~R., Bautista, M., Collins, N.~R., Corcoran, M.~F., Damineli, A., Hamann,
  F., Hartman, H., Johansson, S., Stahl, O., \& Weis, K. 2004, The
  Astrophysical Journal, 605, 405

\bibitem[{Soker(2003)}]{Soker:2003p513}
Soker, N. 2003, ApJ, 597, 513

\bibitem[{Soker(2005)}]{Soker:2005p540}
---. 2005, The Astrophysical Journal, 635, 540

\bibitem[{Soker(2007)}]{Soker:2007p482}
---. 2007, The Astrophysical Journal, 661, 482

\bibitem[{Soker \& Behar(2006)}]{Soker:2006p1563}
Soker, N. \& Behar, E. 2006, The Astrophysical Journal, 652, 1563

\bibitem[{Sota {et~al.}(2011)Sota, Ma{\'\i}z~Apell{\'a}niz, Walborn, Alfaro,
  Barb{\'a}, Morrell, Gamen, \& Arias}]{Sota:2011p24}
Sota, A., Ma{\'\i}z~Apell{\'a}niz, J., Walborn, N.~R., Alfaro, E.~J.,
  Barb{\'a}, R.~H., Morrell, N.~I., Gamen, R.~C., \& Arias, J.~I. 2011, The
  Astrophysical Journal Supplement, 193, 24

\bibitem[{Steiner \& Damineli(2004)}]{Steiner:2004pL133}
Steiner, J.~E. \& Damineli, A. 2004, The Astrophysical Journal, 612, L133

\bibitem[{Teodoro {et~al.}(2008)Teodoro, Damineli, Sharp, Groh, \&
  Barbosa}]{Teodoro:2008p564}
Teodoro, M., Damineli, A., Sharp, R.~G., Groh, J.~H., \& Barbosa, C.~L. 2008,
  Monthly Notices of the Royal Astronomical Society, 387, 564

\bibitem[{Thackeray(1953)}]{Thackeray:1953p211}
Thackeray, A.~D. 1953, Monthly Notices of the Royal Astronomical Society, 113,
  211

\bibitem[{Thackeray(1967)}]{Thackeray:1967p51}
---. 1967, Monthly Notices of the Royal Astronomical Society, 135, 51

\bibitem[{Tody(1993)}]{Tody:1993p173}
Tody, D. 1993, Astronomical Data Analysis Software and Systems II, 52, 173

\bibitem[{Usov(1992)}]{Usov:1992p635}
Usov, V.~V. 1992, ApJS, 389, 635

\bibitem[{van Genderen {et~al.}(2006)van Genderen, Sterken, Allen, \&
  Walker}]{vanGenderen:2006p3}
van Genderen, A.~M., Sterken, C., Allen, W.~H., \& Walker, W. S.~G. 2006,
  Journal of Astronomical Data, 12, 3

\bibitem[{Verner {et~al.}(2005)Verner, Bruhweiler, \& Gull}]{Verner:2005p973}
Verner, E., Bruhweiler, F., \& Gull, T. 2005, ApJ, 624, 973

\bibitem[{Walborn {et~al.}(2010)Walborn, Sota, Ma{\'\i}z~Apell{\'a}niz, Alfaro,
  Morrell, Barba, Arias, \& Gamen}]{Walborn:2010pL143}
Walborn, N.~R., Sota, A., Ma{\'\i}z~Apell{\'a}niz, J., Alfaro, E.~J., Morrell,
  N.~I., Barba, R.~H., Arias, J.~I., \& Gamen, R.~C. 2010, ApJS, 711, L143

\bibitem[{Weigelt {et~al.}(1995)Weigelt, Albrecht, Barbieri, Blades,
  Boksenberg, Crane, Davidson, Deharveng, Disney, Jakobsen, Kamperman, King,
  Macchetto, Mackay, Paresce, Baxter, Greenfield, Jedrzejewski, Nota, \&
  Sparks}]{Weigelt:1995p11}
Weigelt, G., Albrecht, R., Barbieri, C., Blades, J.~C., Boksenberg, A., Crane,
  P., Davidson, K., Deharveng, J.~M., Disney, M.~J., Jakobsen, P., Kamperman,
  T.~M., King, I.~R., Macchetto, F., Mackay, C.~D., Paresce, F., Baxter, D.,
  Greenfield, P., Jedrzejewski, R., Nota, A., \& Sparks, W.~B. 1995, Revista
  Mexicana de Astronomia y Astrofisica Serie de Conferencias, 2, 11

\bibitem[{Weigelt \& Ebersberger(1986)}]{Weigelt:1986pL5}
Weigelt, G. \& Ebersberger, J. 1986, A{\&}A, 163, L5

\bibitem[{White \& Becker(1995)}]{White:1995p352}
White, R.~L. \& Becker, R.~H. 1995, ApJ, 451, 352

\bibitem[{Whitelock {et~al.}(1994)Whitelock, Feast, Koen, Roberts, \&
  Carter}]{Whitelock:1994p364}
Whitelock, P.~A., Feast, M.~W., Koen, C., Roberts, G., \& Carter, B.~S. 1994,
  R.A.S. MONTHLY NOTICES V.270, 270, 364

\bibitem[{Williams(2011)}]{Williams:2011p595}
Williams, P. 2011, Bulletin de la Societe Royale des Sciences de Liege, 80, 595

\bibitem[{Williams {et~al.}(1990)Williams, van~der Hucht, Pollock, Florkowski,
  van~der Woerd, \& Wamsteker}]{Williams:1990p662}
Williams, P.~M., van~der Hucht, K.~A., Pollock, A. M.~T., Florkowski, D.~R.,
  van~der Woerd, H., \& Wamsteker, W.~M. 1990, Monthly Notices of the Royal
  Astronomical Society, 243, 662

\bibitem[{Zanella {et~al.}(1984)Zanella, Wolf, \& Stahl}]{Zanella:1984p79}
Zanella, R., Wolf, B., \& Stahl, O. 1984, A{\&}A, 137, 79

\end{thebibliography}

\clearpage

\LongTables 

\begin{deluxetable}{ccccc}


\tablecaption{Journal of observations and measurements for the low-excitation events \#9, \#10, \#11 and \#12.\label{tbl-1}}

\tablehead{\colhead{\multirow{3}{*}{JD}} & \colhead{cycle} & \colhead{Equivalent} & \colhead{Radial velocity} & \colhead{Observatory} \\
 & + & width & of the peak\tablenotemark{a} & + \\
 & phase & (\AA) & (km\,\,s$^{-1}$) & Instrument }

\startdata

2448776.4 &  9.001 & $-1.27\pm 0.10$ & $-$293 & OPD+Coud\'e \\
2448794.5 &  9.010 & $-0.31\pm 0.10$ & $-$75 & ESO/La Silla+FLASH/HEROS \\
2448825.5 &  9.025 & $-0.56\pm 0.10$ & $-$25 & ESO/La Silla+FLASH/HEROS \\
2448830.5 &  9.028 & $-0.36\pm 0.10$ & $-$10 & ESO/La Silla+FLASH/HEROS \\
2448839.5 &  9.032 & $-0.26\pm 0.10$ & $-$40 & ESO/La Silla+FLASH/HEROS \\
2448844.5 &  9.035 & $-0.19\pm 0.10$ & $-$35 & ESO/La Silla+FLASH/HEROS \\
2450771.1 &  9.987 & $-1.93\pm 0.10$ & $-$240 & MSSSO+Coud\'e \\
2450808.2 & 10.006 & $ 0.20\pm 0.10$ & N.M. & MSSSO+Coud\'e \\
2450869.9 & 10.036 & $-0.19\pm 0.10$ & $-$34 & MSSSO+Coud\'e \\
2450896.1 & 10.049 & $-0.15\pm 0.10$ & $-$10 & MSSSO+Coud\'e \\
2452773.5 & 10.977 & $-0.72\pm 0.12$ &  $-$45 & OPD+Coud\'e \\
2452803.5 & 10.992 & $-1.48\pm 0.10$ & $-$248 & OPD+Coud\'e \\
2452811.4 & 10.996 & $-2.21\pm 0.55$ & $-$266 & OPD+Coud\'e \\
2452812.4 & 10.996 & $-2.67\pm 0.10$ & $-$264 & OPD+Coud\'e \\
2452813.5 & 10.997 & $-2.30\pm 0.12$ & $-$295 & OPD+Coud\'e \\
2452814.4 & 10.997 & $-2.36\pm 0.63$ & $-$349 & OPD+Coud\'e \\
2452815.4 & 10.998 & $-2.80\pm 0.14$ & $-$356 & OPD+Coud\'e \\
2452816.4 & 10.998 & $-2.88\pm 0.13$ & $-$356 & OPD+Coud\'e \\
2452817.4 & 10.999 & $-2.48\pm 0.10$ & $-$368 & OPD+Coud\'e \\
2452818.4 & 10.999 & $-1.93\pm 0.10$ & $-$390 & OPD+Coud\'e \\
2452819.4 & 11.000 & $-1.75\pm 0.14$ & $-$405 & OPD+Coud\'e \\
2452820.5 & 11.000 & $-1.16\pm 0.10$ & $-$432 & OPD+Coud\'e \\
2452821.4 & 11.001 & $-0.48\pm 0.39$ & $-$442 & OPD+Coud\'e \\
2452822.4 & 11.001 & $-0.29\pm 0.11$ & $-$402 & OPD+Coud\'e \\
2452823.4 & 11.002 & $-0.09\pm 0.11$ & $-$460 & OPD+Coud\'e \\
2452825.5 & 11.003 & $ 0.00\pm 0.10$ &  N.M. & OPD+Coud\'e \\
2452826.4 & 11.003 & $-0.08\pm 0.10$ &  N.M. & OPD+Coud\'e \\
2452827.4 & 11.004 & $-0.08\pm 0.10$ &  N.M. & OPD+Coud\'e \\
2452828.4 & 11.004 & $-0.07\pm 0.10$ &  N.M. & OPD+Coud\'e \\
2452830.5 & 11.005 & $-0.09\pm 0.10$ &  N.M. & OPD+Coud\'e \\
2452841.4 & 11.011 & $-0.65\pm 0.16$ &  $-$24 & OPD+Coud\'e \\
2452987.7 & 11.083 & $-0.12\pm 0.17$ &   $-$1 & OPD+Coud\'e \\
2454190.7 & 11.678 & $-0.04\pm 0.12$ &  $-$97 & OPD+Coud\'e \\
2454311.4 & 11.737 & $-0.07\pm 0.13$ &   $-$1 & OPD+Coud\'e \\
2454595.4 & 11.878 & $-0.07\pm 0.12$ &    +5 & OPD+Coud\'e \\
2454596.5 & 11.878 & $-0.11\pm 0.10$ &    +4 & OPD+Coud\'e \\
2454598.4 & 11.879 & $-0.08\pm 0.10$ &   $-$4 & OPD+Coud\'e \\
2454774.5 & 11.966 & $-0.55\pm 0.20$ &  $-$50 & LCO+MIKE \\
2454774.8 & 11.967 & $-0.55\pm 0.20$ & $-$50 & LCO+MIKE\\
2454784.9 & 11.972 & $-0.43\pm 0.20$ & $-$75 & LCO+MIKE\\
2454786.0 & 11.972 & $-0.42\pm 0.05$ &  $-$70 & CASLEO+EBASIM \\
2454787.0 & 11.973 & $-0.34\pm 0.01$ &  $-$75 & CASLEO+EBASIM \\
2454803.8 & 11.981 & $-0.61\pm 0.16$ & $-$101 & SOAR+Goodman \\
2454809.8 & 11.984 & $-1.32\pm 0.14$ & $-$109 & SOAR+Goodman \\
2454815.8 & 11.987 & $-2.19\pm 0.20$ & $-$127 & LCO+MIKE\\
2454815.8 & 11.987 & $-2.30\pm 0.11$ & $-$154 & SOAR+Goodman \\
2454821.7 & 11.990 & $-1.71\pm 0.12$ & $-$175 & SOAR+Goodman \\
2454821.8 & 11.990 & $-1.74\pm 0.11$ & $-$155 & ESO/La Silla+FEROS \\
2454822.9 & 11.990 & $-1.81\pm 0.10$ & $-$153 & ESO/La Silla+FEROS \\
2454823.8 & 11.991 & $-1.78\pm 0.11$ & $-$151 & ESO/La Silla+FEROS \\
2454824.8 & 11.991 & $-1.69\pm 0.11$ & $-$163 & ESO/La Silla+FEROS \\
2454825.8 & 11.992 & $-1.75\pm 0.11$ & $-$156 & ESO/La Silla+FEROS \\
2454826.8 & 11.992 & $-1.65\pm 0.11$ & $-$146 & SOAR+Goodman \\
2454827.8 & 11.993 & $-1.70\pm 0.14$ & $-$195 & SOAR+Goodman \\
2454827.9 & 11.993 & $-1.61\pm 0.11$ & $-$163 & ESO/La Silla+FEROS \\
2454828.8 & 11.993 & $-1.53\pm 0.10$ & $-$166 & SOAR+Goodman \\
2454828.9 & 11.993 & $-1.49\pm 0.12$ & $-$168 & ESO/La Silla+FEROS \\
2454829.8 & 11.994 & $-1.45\pm 0.11$ & $-$187 & SOAR+Goodman \\
2454829.9 & 11.994 & $-1.40\pm 0.11$ & $-$196 & ESO/La Silla+FEROS \\
2454830.8 & 11.994 & $-1.47\pm 0.10$ & $-$260 & SOAR+Goodman \\
2454830.9 & 11.994 & $-1.46\pm 0.10$ & $-$228 & ESO/La Silla+FEROS \\
2454831.8 & 11.995 & $-1.87\pm 0.11$ & $-$255 & SOAR+Goodman \\
2454831.9 & 11.995 & $-1.85\pm 0.12$ & $-$250 & ESO/La Silla+FEROS \\
2454832.9 & 11.995 & $-2.55\pm 0.11$ & $-$224 & ESO/La Silla+FEROS \\
2454833.9 & 11.996 & $-2.77\pm 0.11$ & $-$218 & ESO/La Silla+FEROS \\
2454834.8 & 11.996 & $-2.48\pm 0.11$ & $-$269 & SOAR+Goodman \\
2454834.9 & 11.996 & $-2.54\pm 0.11$ & $-$248 & ESO/La Silla+FEROS \\
2454835.0 & 11.996 & $-2.34\pm 0.01$ & $-$314 & CASLEO+REOSC \\
2454835.8 & 11.997 & $-2.40\pm 0.10$ & $-$319 & SOAR+Goodman \\
2454836.0 & 11.997 & $-2.49\pm 0.01$ & $-$378 & CASLEO+REOSC \\
2454837.0 & 11.997 & $-2.40\pm 0.01$ & $-$362 & CASLEO+REOSC \\
2454837.8 & 11.998 & $-2.57\pm 0.13$ & $-$341 & SOAR+Goodman \\
2454838.0 & 11.998 & $-2.08\pm 0.01$ & $-$336 & CASLEO+REOSC \\
2454839.8 & 11.999 & $-1.90\pm 0.13$ & $-$340 & SOAR+Goodman \\
2454840.0 & 11.999 & $-1.91\pm 0.01$ & $-$349 & CASLEO+REOSC \\
2454841.0 & 11.999 & $-1.79\pm 0.01$ & $-$370 & CASLEO+REOSC \\
2454841.8 & 12.000 & $-1.67\pm 0.12$ & $-$362 & SOAR+Goodman \\
2454842.0 & 12.000 & $-1.87\pm 0.01$ & $-$383 & CASLEO+REOSC \\
2454842.8 & 12.000 & $-1.48\pm 0.10$ & $-$389 & SOAR+Goodman \\
2454843.7 & 12.001 & $-1.21\pm 0.12$ & $-$412 & OPD+Coud\'e \\
2454843.8 & 12.001 & $-0.93\pm 0.10$ & $-$420 & SOAR+Goodman \\
2454844.0 & 12.001 & $-0.65\pm 0.01$ & $-$427 & CASLEO+REOSC \\
2454844.7 & 12.001 & $-0.89\pm 0.11$ & $-$387 & OPD+Coud\'e \\
2454844.8 & 12.001 & $-0.63\pm 0.11$ & $-$424 & SOAR+Goodman \\
2454845.0 & 12.001 & $-0.63\pm 0.01$ & $-$452 & CASLEO+REOSC \\
2454845.8 & 12.002 & $-0.32\pm 0.10$ & $-$393 & SOAR+Goodman \\
2454846.8 & 12.002 & $-0.09\pm 0.11$ & $-$411 & SOAR+Goodman \\
2454847.0 & 12.002 & $-0.04\pm 0.01$ & $-$425 & CASLEO+REOSC \\
2454847.8 & 12.003 & $ 0.04\pm 0.10$ & N.M. & SOAR+Goodman \\
2454848.0 & 12.003 & $ 0.02\pm 0.01$ & N.M. & CASLEO+REOSC \\
2454848.8 & 12.003 & $-0.00\pm 0.20$ & N.M. & LCO+MIKE\\
2454849.0 & 12.003 & $-0.09\pm 0.01$ & N.M. & CASLEO+REOSC \\
2454850.0 & 12.004 & $ 0.08\pm 0.01$ & N.M. & CASLEO+REOSC \\
2454852.0 & 12.005 & $-0.05\pm 0.01$ & N.M. & CASLEO+REOSC \\
2454853.0 & 12.005 & $-0.02\pm 0.01$ & N.M. & CASLEO+REOSC \\
2454853.7 & 12.006 & $-0.13\pm 0.11$ & N.M. & SOAR+Goodman \\
2454854.0 & 12.006 & $-0.12\pm 0.11$ & N.M. & CASLEO+REOSC \\
2454854.8 & 12.006 & $+0.01\pm 0.20$ & N.M. & LCO+MIKE\\
2454855.0 & 12.006 & $-0.07\pm 0.11$ & N.M. & CASLEO+REOSC \\
2454859.8 & 12.009 & $-0.42\pm 0.11$ & $-$90 & SOAR+Goodman \\
2454871.5 & 12.014 & $-0.76\pm 0.20$ & $-$50 & LCO+MIKE\\
2454871.7 & 12.014 & $-0.70\pm 0.11$ & $-$209 & SOAR+Goodman \\
2454882.6 & 12.020 & $-1.18\pm 0.12$ & $-$15 & OPD+Coud\'e \\
2454883.7 & 12.021 & $-1.13\pm 0.20$ & 0. & LCO+MIKE\\
2454888.5 & 12.023 & $-1.03\pm 0.12$ &    +6 & SOAR+Goodman \\
2454898.0 & 12.027 & $-0.11\pm 0.08$ &    0 & CASLEO+EBASIM \\
2454899.5 & 12.028 & $-0.24\pm 0.11$ &    +3 & SOAR+Goodman \\
2454905.0 & 12.031 & $-0.40\pm 0.33$ &    0 & CASLEO+REOSC \\
2454906.0 & 12.031 & $-0.20\pm 0.07$ &    0 & CASLEO+REOSC \\
2454907.0 & 12.032 & $-0.18\pm 0.32$ &    0 & CASLEO+REOSC \\
2454925.5 & 12.041 & $-0.10\pm 0.11$ &    +7 & SOAR+Goodman \\
2454940.5 & 12.048 & $-0.32\pm 0.11$ &    +0 & OPD+Coud\'e \\
2454942.5 & 12.049 & $-0.07\pm 0.11$ &   $-$9 & SOAR+Goodman \\
2454953.5 & 12.055 & $-0.06\pm 0.11$ &  $-$21 & SOAR+Goodman \\
2454953.6 & 12.055 & $-0.00\pm 0.11$ &   +13 & ESO/La Silla+FEROS \\
2454955.6 & 12.056 & $-0.01\pm 0.11$ &   +13 & ESO/La Silla+FEROS \\
2454967.5 & 12.062 & $ 0.02\pm 0.11$ &   +14 & SOAR+Goodman\\
2455368.4 & 12.260 & $-0.05\pm 0.14$ &  $-$100 & OPD+Coud\'e \\
2455754.8 & 12.451 & $-0.10\pm 0.10$ & +30 &  OPD+Coud\'e

\enddata

\tablenotetext{(a)}{Velocity of the highest point of the fitting of the \ion{He}{2}~$\lambda4686$ line profile. In this table, the flag N.M. (\emph{not measurable}) indicates when the line is absent (or very weak) and 
when the peak velocity is impossible to measure.}

\end{deluxetable}

\end{document}